%% file: SNe.tex
\documentclass[fleqn, usenatbib]{mnras}
\usepackage{mathptmx}

\usepackage[T1]{fontenc}
\usepackage{ae,aecompl}

\usepackage{amssymb}
\usepackage{amsmath}
\usepackage{mathtools}
\usepackage{appendix}

\usepackage[usenames, dvipsnames]{color}
\usepackage[normalem]{ulem}

\title[Clustered Supernovae]{Enhanced Momentum Feedback from Clustered Supernovae}

\author[E. S. Gentry et al.]{
Eric S. Gentry,$^{1}$\thanks{E-mail: egentry@ucsc.edu}
Mark R. Krumholz,$^{2}$
Avishai Dekel$^{3}$
and Piero Madau$^{1,4}$
\\
$^{1}$Department of Astronomy and Astrophysics, University of California at Santa Cruz, 1156 High St., Santa Cruz, CA, 95064, USA\\
$^{2}$Research School of Astronomy \& Astrophysics, Australian National University, Canberra, ACT 2611, Australia\\
$^{3}$Center for Astrophysics and Planetary Science, Racah Institute of Physics, The Hebrew University, Jerusalem 91904, Israel\\
$^{4}$Institut d'Astrophysique de Paris, Sorbonne Universit{\'e}s, UPMC Univ Paris 6 et CNRS, UMR 7095, 98 bis bd Arago, 75014 Paris, France
}

\date{Accepted 2016 October 20. Received 2016 October; in original form 2016 June 02}

\pubyear{2016}

\begin{document}
\label{firstpage}
\pagerange{\pageref{firstpage}--\pageref{lastpage}}
\maketitle

\begin{abstract} 
Young stars typically form in star clusters, so the supernovae (SNe) they produce are clustered in space and time.  This clustering of SNe may alter the momentum per SN deposited in the interstellar medium (ISM) by affecting the local ISM density, which in turn affects the cooling rate.
We study the effect of multiple SNe using idealized 1D hydrodynamic simulations which explore a large parameter space of the number of SNe, and the background gas density and metallicity.  The results are provided as a table and an analytic fitting formula.  We find that for clusters with up to $\sim 100$ SNe the asymptotic momentum scales super-linearly with the number of SNe, resulting in a momentum per SN that can be an order of magnitude larger than for a single SN, with a maximum efficiency for clusters with $10-100$ SNe.
We argue that additional physical processes not included in our simulations -- self-gravity, breakout from a galactic disk, and galactic shear -- can slightly reduce the momentum enhancement from clustering, but the average momentum per SN still remains a factor of 4 larger than the isolated SN value when averaged over a realistic cluster mass function for a star-forming galaxy. 
We conclude with a discussion of the possible role of mixing between hot and cold gas, induced by multi-dimensional instabilities or preexisting density variations, as a limiting factor in the buildup of momentum by clustered SNe, and suggest future numerical experiments to explore these effects.
\end{abstract}



\section{Introduction}
\label{section:intro}

Supernovae (SNe) play a key role in regulating star formation at galactic scales.  SN energy, if retained, can disrupt molecular clouds and small galaxies \citep{1986ApJ...303...39D}.  Even if significant energy is lost to radiative cooling, SNe inject momentum that cannot be radiated away, which drives turbulence, the dominant form of dynamical pressure support in galactic discs \citep{2011ApJ...743...25K,2011ApJ...734...65J}. This turbulent support both prevents the collapse of star-forming regions  \citep[locally limiting star formation;][]{2011ApJ...731...41O,2013MNRAS.433.1970F}, and drives galactic winds \citep[globally limiting star formation;][]{2005ApJ...618..569M,2012MNRAS.421.3522H,2013MNRAS.432..455D,2013MNRAS.429.1922C,2016MNRAS.455..334T}.

Unfortunately, the processes controlling supernova remnant (SNR) evolution operate at smaller scales than what can typically be resolved by galaxy or cosmological simulations.  In particular, the dense shells of SNRs rapidly radiate away most of the SN energy, leaving a cold dense shell and a hot diffuse interior. If a simulation cannot resolve these two zones, then it cannot realistically evolve the SNR, resulting in problems such as over-cooling \citep{1992ApJ...391..502K}. To counteract this, some authors have prescribed turning off cooling for young, unresolved SNRs \citep{1997PhDT........19G, 2006MNRAS.373.1074S}, while others have proposed models which mimic the otherwise unresolved multi-phase nature of the interstellar medium \citep[ISM;][]{2014MNRAS.442.3013K}.  These methods have their strengths, but the most direct way to incorporate the relevant physics is multiscale modeling: evolve a number of SNRs in a representative set of environments using simulations with high enough resolution to resolve the relevant processes, and use those results in large, low resolution simulations.

Early attempts at using high resolution simulations to create subgrid SN feedback models focused on producing energy-driven models \citep{1998ApJ...500...95T}, but recently there has been increased interest in momentum-driven models, both by those using high resolution simulations to study SNRs directly \citep{2015MNRAS.450..504M,2015MNRAS.451.2757W,2015ApJ...802...99K,2015A&A...576A..95I} and by those who use such models in lower resolution simulations \citep{2011MNRAS.417..950H,2012ApJ...754....2S,2014A&A...570A..81H,2016ApJ...827...28G}.  This change in emphasis has been driven by the realization that, while the energy content of SNRs is important for producing hot galactic winds that are observable in X-rays, the momentum budget is more important when it comes to SNRs regulating star formation and possibly ejecting cool gas from galaxies \citep{2013MNRAS.432..455D}.

At early times, before radiative losses are important, a SNR is in the \emph{Sedov} stage, during which the energy is approximately conserved and the radial momentum is increasing.  Once radiative losses become significant, it enters a \emph{pressure-driven snowplow} phase, during which the energy is decreasing and the momentum is still increasing.  As the bubble expands and cools (adiabatically and radiatively), its pressure will eventually decrease to the ISM pressure, at which point it becomes a \emph{momentum-driven snowplow}.  Asymptotically, in the idealized case of a spherical SNR expanding into a uniform, cold medium, this results in zero energy being added to the ISM, but a non-zero and finite amount of momentum being added. The goal of high resolution simulations is to follow all of these phases, and identify the asymptotic momentum as a function of the properties of the driving stars and the large-scale environment, making this value available for use in larger-scale models.

A number of authors have performed systematic parameter studies of the expansion of a SNR from a single SN in spherical symmetry \citep{1974ApJ...188..501C,1988ApJ...334..252C,1998ApJ...500...95T}. The most complete of these studies, that of \citet{1998ApJ...500...95T}, spanned metallicities from $10^{-3}-10^{0.5}$ times Solar and ambient densities from $0.1-10^3$ H atoms cm$^{-3}$. More recently, there have been a number of 3D simulations which allowed the study of SNR evolution within a more realistic, non-spherically symmetric background. \citet{2015MNRAS.450..504M}, \citet{2015ApJ...802...99K}, and \citet{2015MNRAS.451.2757W} all found that inhomogeneities present prior to the first SN explosion -- such as those expected due to a multi-phase structure of the ISM or ionized bubbles created by pre-SN radiation -- did not change the final momentum by more than 60\%. A more interesting effect was found by considering the inhomogeneities that result from bubbles of previous SNRs.  \cite{2015ApJ...802...99K} found that a series of clustered SNe can \emph{decrease} the momentum per SN, in some cases by almost a factor of two.  On the other hand, \cite{2015MNRAS.451.2757W} found that multiple SNe might \emph{increase} the momentum per SNe, by at least 25\%, depending on the delay time between SNe. The dependence on delay time further complicates this discrepancy between authors since neither set of authors used realistic delay time distributions for the number of SNe considered.  \citet{2016arXiv160300815Y} used 3D simulations to study how clustered SNe can merge into a superbubble (using a realistic SN delay time distribution), but did not study the momentum produced and did not test the effect of gas metallicity. So we are left with a series of questions: for a realistic delay time distribution of clustered SNe, does the momentum per SN increase or decrease relative to single SN models, and by how much?  Does the result depend on the density or metallicity of the environment in which the SNe explode?  Does it vary systematically with the number of SNe that are clustered together?

In this paper we seek to measure directly the impact that clustering has on the momentum budget of SNe.  In order to sample a wide range of densities, metallicities and cluster sizes, we create a suite of several hundred 1D, spherically-symmetric simulations.  Using a 1D geometry means we lose the ability to simulate non-spherically symmetric inhomogeneities but in doing so we gain the ability to probe a far wider parameter space than any previous studies of multiple SNe, and to achieve far higher spatial resolutions than previous works studying momentum feedback. As we will show, both are necessary for understanding how clustering impacts momentum feedback.

The remainder of this paper is as follows. In \autoref{section:methods} we discuss the numerical methods used in this study. The numerical results of our simulations are presented in \autoref{section:results}.  In \autoref{section:momentum} we use these results to build a model that can predict the momentum injection per SN as a function of density, metallicity and number of SNe, in a form suitable for inclusion in subgrid and analytic models. We discuss the significance of our results and model in \autoref{section:discussion}, comparing to previous works.  Finally, we summarize our conclusions in \autoref{section:conclusion}.


\section{Numerical Methods}
\label{section:methods}
Our simulations make use of a custom-built 1D spherically-symmetric moving-mesh code that solves the finite volume equations of compressible hydrodynamics.  Our code includes radiative cooling and injection of mass and energy by both SNe and pre-SN winds. The star cluster is assumed to lie at the centre of our simulation, with our computational domain beginning just outside the cluster and extending outwards. SN ejecta and pre-SN winds are added to the innermost zone of this domain.  We run these simulations until all SNe have occurred and the momentum reaches a maximum.

In the rest of this section we go into greater depth on the numerical methods used in our simulations and the limitations of our assumptions.  In \hyperref[section:verification]{Appendix~\ref*{section:verification}} we test our code against both the analytic Sedov solution and the earlier numerical results of \citet{1998ApJ...500...95T} for isolated SNe, and verify that it reproduces them well. For the interested reader, our code has been publicly released\footnote{Source code available at: \url{github.com/egentry/clustered_SNe}}.

\subsection{Initial Conditions}
\label{section:methods:initial}
All our simulations begin with a star cluster of mass $M_\mathrm{cluster}$ placed at the origin surrounded by an initially-uniform, stationary ideal gas with adiabatic index $\gamma = 5/3$.  We vary $M_\mathrm{cluster}$ from $10^2 - 10^5$ $ M_\odot$ (using steps of 1 dex, with an additional step at $10^{2.5} $ $M_\odot$ to better resolve a key region of parameter space). We explore gas mass densities in steps of 1 dex, ranging from $\rho_0 = 1.33 \times 10^{-3} - 10^{2}$ $m_\mathrm{H}$ cm$^{-3}$, where $m_\mathrm{H} = 1.67 \times 10^{-24}$ g is the mass of the hydrogen atom, corresponding to gas number densities of $n_0 = \rho_0 / (1.33 m_\mathrm{H})$ H nuclei cm$^{-3}$ for a helium mass fraction $Y=0.23$ and a metal mass fraction $Z=.02$.	  We consider gas metallicities in steps of half a dex, ranging from $Z_0 = 10^{-3} - 10^{0.5}Z_\odot$, excluding $10^{-2.5} Z_\odot$, where we have taken solar metallicity to be $Z_\odot = 0.02$. 
The density and metallicity grids are chosen to closely match those explored by \cite{1998ApJ...500...95T}. Consistent with \citeauthor{1998ApJ...500...95T}, we compute mean molecular weights assuming a fixed helium mass fraction, $Y=0.23$, taking the remainder to be hydrogen ($X = 1 - Z - Y$). The gas has an initial temperature of $10^4$ K, but is allowed to cool to lower temperatures via radiation -- see \autoref{section:methods:cooling}.

Our simulations start with 1024 zones, linearly spaced from radii 
$R_\mathrm{in}$ to $R_\mathrm{out}$, which follow the scaling
\begin{align}
	R_\mathrm{out} &= 300 
	\left( \frac{\rho_0}{1.33 m_\mathrm{H} \text{ cm}^{-3} } \right)^{-1/3}
	\left( \frac{M_\mathrm{cluster}}{100 M_\odot } \right)^{1/3}
	\mathrm{pc} \label{eq:R:out}
	\\
	R_\mathrm{in} &= 10^{-4} R_\mathrm{out} \label{eq:R:in}
\end{align}
This scaling is somewhat arbitrary and was set by initial tests; it was chosen to approximately reflect the final size of each simulation.
If the outer boundary is too small, the domain will automatically extend
when a shock nears the outer boundary.

\subsection{Hydrodynamics}
\label{section:methods:hydro}

Our code solves the equations of compressible hydrodynamics in spherical symmetry using a moving-mesh finite volume method, including source terms for radiative cooling.  Our method is an extension of the one implemented by \citet{2016ApJ...821...76D}. The equations we solve are
\begin{equation}
	\frac{ \mathrm{d} }{ \mathrm{d}t} \int \boldsymbol{U} \mathrm{d}V - \int \boldsymbol{F} \mathrm{d}A = \boldsymbol{S}
\end{equation}
where $\boldsymbol{U} \mathrm{d} V$ is the vector of conserved quantities
\begin{equation}
	\boldsymbol{U} = \begin{pmatrix} \begin{array}{l}
						\rho \\ 
						\rho u_r \\ 
						\rho e  \\
						\rho Z
					\end{array} \end{pmatrix}
\end{equation}
for a density $\rho$, a bulk fluid velocity $u$,
a specific total energy $e$ and a local metallicity $Z$. The quantity
$\boldsymbol{F}$ is the conservative flux, given by
\begin{equation}
	\boldsymbol{F} = \begin{pmatrix} \begin{array}{lll}
						(u_r - w_r) \rho                  \\ 
						(u_r - w_r) \rho u_r & + P         \\ 
						(u_r - w_r) \rho e   & + P u_r  + H \\
						(u_r - w_r) \rho Z                   \\ 

					\end{array} \end{pmatrix}
\end{equation}
where $w$ is the computational mesh velocity, which is set to be the average velocity of the two zones adjacent to the boundary
\begin{equation}
  	w_r^{(i+1/2)} = \frac{u_r^{(i)} + u_r^{(i+1)}}{2}
\end{equation} 
approximating Lagrangian hydrodynamics. Here $P$ is the pressure given by
\begin{equation}
	P = (\gamma - 1) \rho e_\mathrm{int}  
\end{equation}
and $e_\mathrm{int}$ is the specific internal energy:
\begin{equation}
e_\mathrm{int} = e - \frac{1}{2} v_r^2 . \label{eq:energy:internal:indirect}
\end{equation}
At the inner boundary, we enforce a zero flux boundary condition; the outer boundary condition does not matter, as we automatically add zones before the shock reaches the outer boundary (and we have assumed the background is homogeneous).

This formulation implicitly introduces artificial mesh viscosity, particular at the inner boundary (i.e.\ wall heating), which leads to unphysically high temperatures. We counteract this by explicitly including an artificial conduction term, $H$. We use the artificial conduction prescription of \citet[Eq.\ 2.3]{Noh:1987:ECS:33108.33111}:
\begin{equation}
	H = 
		\begin{cases}
		h_0 \rho |\Delta u| \Delta e_\mathrm{int} + h_1 \rho c_\mathrm{s} \Delta e_\mathrm{int} & \Delta u < 0
		\\
		0 & \Delta u > 0
		\end{cases}	
	\label{eq:conduction}
\end{equation}
where $c_s$ is the adiabatic sound speed, $h_0$ and $h_1$ are
tunable constants, typically of order unity, and $\Delta$ represents the differential of a variable across adjacent zones. We chose these constants to be $ h_0 = 0$ and $h_1 = 0.1$, which experimentation showed were the smallest values that were still sufficient to remove most unphysical wall heating. This parameterization is similar to physical conduction in the strong-shock regime with a saturated conduction coefficient which has been lowered by a factor of a few by turbulence and magnetic fields \citep{1977ApJ...211..135C}.

In addition to these conservative fluxes, we also include non-conservative
source terms
\begin{equation}
	\boldsymbol{S} =  \boldsymbol{S}_\mathrm{hydro} 
	 				+ \boldsymbol{S}_\mathrm{cooling}
					+ \boldsymbol{S}_\mathrm{winds}
\end{equation}
\begin{equation}
	\boldsymbol{S}_\mathrm{hydro} = \begin{pmatrix} 
						0 \\ 
						\int 2 (P/r) dV \\ 
						0 \\
						0
					\end{pmatrix}
\end{equation}
\begin{equation}
	\boldsymbol{S}_\mathrm{cooling} = \begin{pmatrix} 
						0 \\ 
						0 \\
						\dot{E}_\mathrm{cooling} \\
						0
					\end{pmatrix}
\end{equation}
\begin{equation}
	\boldsymbol{S}_\mathrm{winds} = \dot{M}_\mathrm{winds} \Delta t \begin{pmatrix} \begin{array}{ll}
						1\\ 
						u_\mathrm{winds} \\ 
						(1/2)u^2_\mathrm{winds} +  e_\text{int, winds}\\
						Z_\mathrm{winds}
					\end{array} \end{pmatrix} .
\end{equation}
We defer a discussion of the cooling rate $\dot{E}_\mathrm{cooling}$ to \autoref{section:methods:cooling}, and a discussion of the wind source term (which is only added to the innermost zone) to \autoref{section:methods:winds}.

The conservative fluxes (excepting conduction) were solved using an HLLC Riemann solver \citep{HLLC} taken from the 
implementation of \citet{2016ApJ...821...76D}.
Artificial conduction and the non-conservative source terms
were handled by operator splitting, solving each term individually.

By using a moving mesh with $w_r \approx u_r$, we approximate Lagrangian hydrodynamics. This reduced numerical errors in the advective flux terms, as well as automatically adjusting to give higher resolution at locations with higher densities (assuming we start with a grid of uniform density).  This improves our accuracy at shocks, where high densities lead to rapid cooling, which drives the subsequent evolution of the SNR.  By using an approximately Lagrangian scheme, we can better resolve the dynamically important regions, without wasting computational time on the less important diffuse bubble.

For strong shocks we need to set a limit on how much zones can expand or compress.  The innermost zone (where SN energy is injected) will significantly expand, so we need to split it in order to retain accuracy where our blasts are being injected.  For computational efficiency we also need to allow zones to merge, because otherwise the zones near the shock become so thin that the computational cost of evolving them is prohibitive. We handle zone splitting and merging using the adaptive mesh algorithm implemented by \citet{2016ApJ...821...76D}: zones thicker (thinner) than $10$ ($0.1$) times the average zone thickness are split (merged).

To improve numerical stability in regions of highly-supersonic flow, we also implement a dual energy formalism.  This approach counteracts the common problem in conservative, total energy codes such as ours that, at Mach numbers greater than unity, the internal energy $e_\mathrm{int}$ can be much smaller than the total energy $e$, so that small truncation errors in $e$ can correspond to an order unity or larger error in $e_\mathrm{int}$, and thus in the temperature and radiative cooling rate. Our dual energy approach is as follows. For most zones and time steps, we follow the update procedure described above and derive $e_\mathrm{int}$ from the mass, momentum and total energy (\autoref{eq:energy:internal:indirect}).  However, in any zone and time step where this procedure yields a value of $e_\mathrm{int} <0$, we instead compute the internal energy via
\begin{equation}
	e_\mathrm{int}(t + \Delta t) = e_\mathrm{int}(t) \left( \frac{dV(t+\Delta t)}{dV(t)}\right)^{1-\gamma} + \Delta e_\mathrm{cool} .
	\label{eq:dualenergy}
\end{equation}
This includes adiabatic heating/cooling and radiative cooling; this ignores advective fluxes, which should be minimal for a pseudo-Lagrangian code, and conductive fluxes. The errors introduced by this dual energy formalism have negligible effects on the overall dynamics and numerical conservation of energy.

\subsection{Cooling}
\label{section:methods:cooling}

Cooling plays a significant role in SNR evolution, with most of the energy from the SN being radiated from a thin, dense shell. To include this cooling, we use the \texttt{Grackle} chemistry and cooling library \citep{2014ApJS..211...19B,2014ApJS..210...14K}, using operator splitting to evolve the thermal energy over each time step.  \texttt{Grackle} sub-steps the thermal evolution using cooling rates pre-computed using \texttt{Cloudy} \citep{1998PASP..110..761F},  assuming ionization equilibrium but not thermal equilibrium between metallicity-dependent optically thin cooling and a cosmological UV background at redshift $z=0$ providing photo-heating and photo-ionization \citep{2012ApJ...746..125H}.
For simplicity we only include heating from a cosmological background, rather than including galactic heating sources.  We leave testing more realistic heating backgrounds and non-ionization equilibrium cooling models for a later work.

\subsection{Cluster Model}
\label{section:methods:cluster}

In order to test the SN momentum produced by a cluster, we need to determine the number of SNe from a cluster and when those SNe will occur.
For each simulation with a given cluster mass, we use the \texttt{SLUG2} code \citep{2012ApJ...745..145D,2014MNRAS.444.3275D,2015MNRAS.452.1447K} to draw the desired mass in stars from a \cite{Kroupa04012002} IMF, using the default ``Stop-nearest'' policy. All stars above an initial mass of $8 $ $M_\odot$ are assumed to result in core-collapse SNe, after stellar lifetimes determined by the Geneva stellar evolution tracks assuming solar metallicity \citep[$Z = 0.014$;][]{2012A&A...537A.146E}. Generally, we find 1 SN per roughly 100 $M_\odot$ of stars, and those SNe occur roughly $3-40$ Myr after the birth of the cluster. Given the power-law tail of the IMF, we expect most SN to come from relatively low mass stars, $M_\star \approx 8 $ $M_\odot$. We do not include Type Ia SNe for most of our simulations, but we do test the impact of short-delay Type Ia SNe in \autoref{section:discussion:additional:Ia}.

Since we are stochastically drawing an IMF, our results for low mass clusters can depend significantly on the random seed.  To minimize the uncertainty in our results induced by this stochasticity, we ran multiple realizations of the lowest cluster masses. Specifically, we ran 9 realizations of each $10^2 $ $M_\odot$ cluster and 4 realizations of each $10^{2.5}$ $ M_\odot$ cluster.

This cluster model is not perfect. The stellar evolution tracks assume a single stellar metallicity, while ideally we would like the tracks to depend on the background metallicity for each simulation. Our cluster model also ignores the effects of stellar rotation and binarity. All of these can affect stellar lifetimes.

\subsection{SN Injection Model}
\label{section:methods:SN}

When a SN occurs, we add energy, mass and metals to the innermost computational zone. For the energy, we adopt a fixed injection of $10^{51}$ ergs per SN. For the mass and metallicity, we use the data of \cite{2007PhR...442..269W}, who provide a grid of SN mass and metal yields as a function of initial stellar mass over a range of initial masses from 12 $M_\odot$ to 120 $M_\odot$. Within this range we linearly interpolate as a function of initial mass; outside this range, we use the nearest neighbor (i.e.\ stars with masses $8-12$ $M_\odot$ are assumed to produce the same yield as 12 $M_\odot$ stars).

As with our cluster model, this SN model is imperfect. 
First, this model over-predicts the ejecta mass for stars with initial masses of $8-12$ $M_\odot$ (the most common progenitors). For example this model predicts that a 9 $M_\odot$ star will eject 9.4 $M_\odot$ of material. This is clearly unphysical, but the true ejected mass is comparable; \citet{2016ApJ...821...38S} show that the true ejected mass ($\approx 7.4$ $M_\odot$) differs by less than 50\% from our simplified model. Overall, this will tend to over-predict the ejecta mass, biasing our results towards slightly more efficient cooling and slightly lower momenta.

Biasing our results in the opposite direction (for fixed cluster mass), we assume all of our massive stars explode, even though some low mass progenitors ($8-9$ $M_\odot$) may not explode \citep{2015ApJ...810...34W} and some high mass progenitors will collapse directly into black holes \citep{2016ApJ...818..124E,2016ApJ...821...38S}, a pathway which can depend significantly on the stellar metallicity \citep{2015ApJ...801...90P}. However, while models differ as to whether more massive stars explode, almost all models agree that all $9-12$ $M_\odot$ stars should explode, so the total number of SNe should not change drastically.

More significantly, SN energies can vary with initial progenitor mass.  In particular, \citet{2016ApJ...821...38S} find that stars with initial masses of $9-12$ $M_\odot$ explode with $<0.7 \times 10^{51}$ ergs of energy.  As these are the most common progenitors, this leads to an IMF-averaged explosion energy of $\approx 0.6 - 0.8 \times 10^{51}$ ergs, depending on the explosion model.  While it is common to assume an explosion energy of $10^{51}$ ergs \citep[e.g.][]{1998ApJ...500...95T,2014ApJS..210...14K,2015ApJ...802...99K,2015A&A...576A..95I,2015MNRAS.450..504M,2015MNRAS.451.2757W}, in doing so we over-predict the average SN energy by a factor of $1.2-1.7$.

As with our stellar evolution tracks, the data of \citet{2007PhR...442..269W} are only for stars of solar metallicity, so we are unable to vary our SN model with the background metallicity. Moreover, we are combining the ejecta computed by \citet{2007PhR...442..269W} with the lifetimes computed by \cite{2012A&A...537A.146E}; these make different assumptions about stellar evolution, and are not fully consistent. Theoretical uncertainties in stellar lifetimes are not too worrisome though; for low mass clusters, we are dominated by stochastic scatter in the IMF; for high mass clusters, the SNe are effectively a continuous wind. The models of \citet{2007PhR...442..269W} and \cite{2012A&A...537A.146E} also differ in the details of pre-SN mass loss, but these discrepancies primarily exist for the most massive stars, which are the least common in our simulations.

\subsubsection{Winds}
\label{section:methods:winds}

If SN ejecta mass is important because more mass leads to faster cooling, then we cannot simply ignore pre-SN mass loss; that mass has to go somewhere.  The pre-SNe mass loss can be determined from the data of \citet{2007PhR...442..269W}, but that does not tell us when that mass was lost, or its physical properties when it was lost (e.g.\ metallicity, wind velocity, wind energy).  For simplicity, we assume that pre-SN mass loss occurs uniformly through a star's lifetime, as a wind with metallicity equal to the background metallicity, at a velocity of $10^3$ km s$^{-1}$, and a temperature of $10^4$ K.  The total mass, metal mass, momentum and energy of this wind are added to the innermost zone.

\section{Numerical Results}
\label{section:results}

\begin{figure}
\includegraphics[width=\columnwidth]{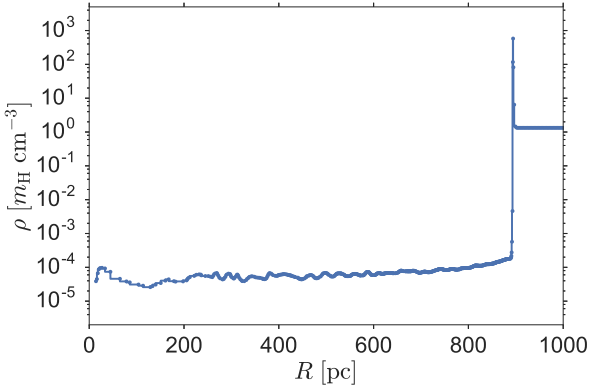}
\caption{
Example density profile of a simulation with $Z=Z_\odot$, $\rho = 1.33$ $ m_\mathrm{H}$ cm$^{-3}$ and $M_\mathrm{cluster} = 10^5$ $M_\odot$ ($N_\mathrm{SNe} = 1008$), shortly after the last SN ($t=38$ Myr).
}
\label{fig:structure:many:lastSN}
\end{figure}

\begin{figure}
\includegraphics[width=\columnwidth]{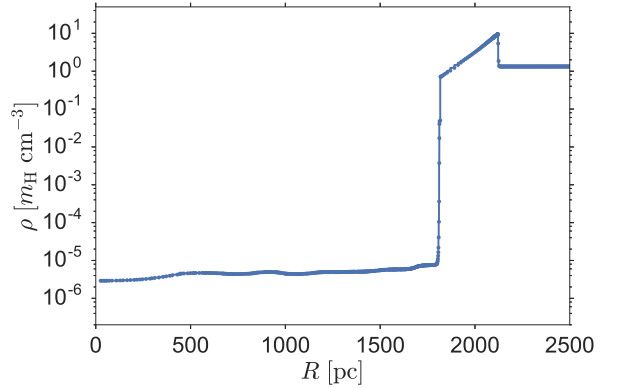}
\caption{
The same simulation as the one shown in \autoref{fig:structure:many:lastSN}, except now at the moment of peak momentum ($t=285$ Myr).  The shock has weakened, causing it to thicken considerably.
}
\label{fig:structure:many:peakmomentum}
\end{figure}

\begin{figure}
\includegraphics[width=\columnwidth]{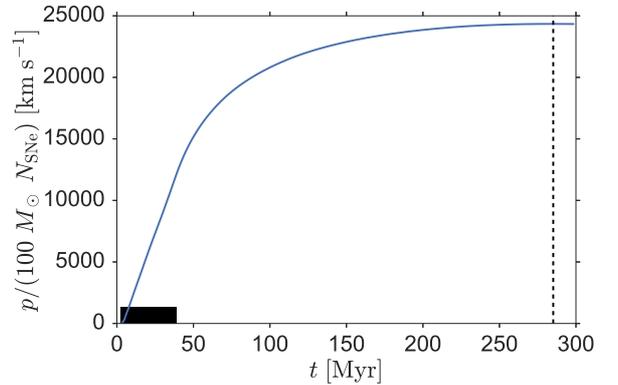}
\caption{
The evolution of the momentum per SN of the cluster shown in \autoref{fig:structure:many:lastSN} and \autoref{fig:structure:many:peakmomentum}.  The time of maximal momentum is marked by a vertical black dashed line; the duration of SN events is denoted by solid black ticks.  For many SNe, the energy injection behaves more like a continuous wind rather than discrete explosions.
}
\label{fig:momentum:evolution:many}
\end{figure}

\begin{figure}
\includegraphics[width=\columnwidth]{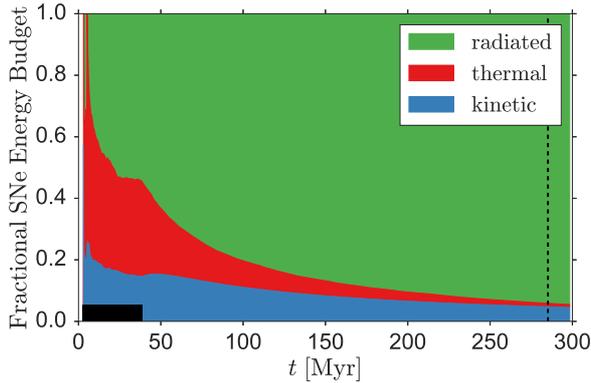}
\caption{
The evolution of the cumulative energy budget of the cluster shown in Figures \ref{fig:structure:many:lastSN} through \ref{fig:momentum:evolution:many}. The time of maximal momentum is marked by a vertical black dashed line; SNe times are denoted by solid black ticks. The kinetic component is measured directly from the simulation; the radiated component is inferred by comparing the decrease in total energy compared to the decrease that would have occurred if there were no SNe and the background cooled anyway; the thermal component is assumed to be whatever remains.
}
\label{fig:energy:evolution:1000}
\end{figure}

\input{results.stub.table.tex}

\input{evolution.stub.table.tex}

We run a total of 672 simulations, sampling a three-dimensional parameter space composed of density $\rho$, metallicity, $Z$ and cluster mass $M_\mathrm{cluster}$. 
As an example of the outcome of our simulations, in \autoref{fig:structure:many:lastSN} we show the density profile immediately after the last SN occurs in the simulation with $\rho = 1.33$ $m_\mathrm{H}$ cm$^{-3}$, $Z=Z_\odot$ and $M_\mathrm{cluster} = 10^5 $ $M_\odot$. In \autoref{fig:structure:many:peakmomentum} we show the density profile for this simulation at the time when the radial momentum reaches its maximum, and we show the momentum as a function of time in \autoref{fig:momentum:evolution:many} and the cumulative energy budget as a function of time in \autoref{fig:energy:evolution:1000}.

In \autoref{tab:results} we provide an overview of our results, extracting the following key parameters when all SNe have occurred and the momentum reaches a maximum: 
the peak momentum $p$; the time, $t$, at which the momentum reaches a maximum (defining $t=0$ as the time of cluster formation); the radius of the shock, $R$, at this time (defined by the furthest zone with an over-density compared to the background); the mass of the remnant, $M_\mathrm{R}$, enclosed by the shock radius at this time; and the kinetic and internal energies, $E_{\mathrm{R,kin}}$ and $E_{\mathrm{R,int}}$, enclosed by the shock radius at this time; finally, we also include a flag for untrustworthy results, which we explain in the next paragraph. 
In \autoref{tab:evolution}, we provide the time-dependent evolution of these parameters for every simulation and each snapshot before the time of peak momentum.

Not all of our simulations are trustworthy.  In some runs, a strong reverse shock reaches the inner boundary before the SNR momentum peaks. In these simulations the shock reflects off our hard inner boundary, whereas in reality the shock converging on the origin would certainly become unstable and would not reflect.  In these cases, we cannot reasonably measure a maximum momentum.  Fortunately, this behavior only occurs in a small part of our parameter space (40/672 runs), and in what follows we will exclude these runs from our analysis.  We also exclude any other realizations of the same initial conditions (an additional 99/672 runs) so as not to bias ourselves by only allowing atypical realizations.  In \autoref{section:momentum:qualitative:few} we explain the astrophysical causes and implications of these flagged runs.

The quantities $M_\mathrm{R}$ and $E_\mathrm{R,int}$ need to be interpreted with some care.  At late times (when these quantities are extracted), the shock has weakened and is becoming a linear sound wave moving through a uniform medium, as illustrated in \autoref{fig:structure:many:peakmomentum}. At this time, the bubble-shell decomposition no longer is a good description, and $M_\mathrm{R}$ and $E_\mathrm{R,int}$ are becoming increasingly dominated by background material which has simply had a sound wave pass through it, but has not been irreversibly affected by a shock.  It would be inaccurate to include this material and its internal energy in SN-driven ``feedback,'' and it is difficult to meaningfully disentangle SN-dominated material and background-dominated material as the SNR is merging into the ISM.  It is easier to disentangle kinetic variables since the background is static; all momentum and kinetic energy must be a result of the SNe.  The kinetic energy $E_\mathrm{R,kin}$ does not asymptote, but it varies slowly at late times, as illustrated in \autoref{fig:energy:evolution:1000}.

\begin{figure}
\includegraphics[width=\columnwidth]{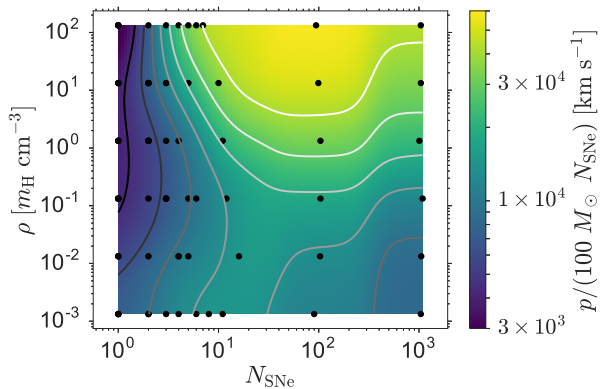}
\caption{
An overview of the final momentum per SN and how it varies with the number of SNe and gas density at fixed metallicity ($Z = Z_\odot$). The locations of our simulations in parameter space are marked by black scatter points (excluding flagged runs), which are not a perfect grid because the numbers of SNe are drawn stochastically. The color image is an interpolation of our simulation results using a Gaussian radial basis function, evaluated on a 100 by 100 grid with 7 greyscale contours logarithmically spaced between $3 \times 10^3$ and $6 \times 10^4$ km s$^{-1}$ (exclusive).
}
\label{fig:momentum:contour:density}
\end{figure}

\begin{figure}
\includegraphics[width=\columnwidth]{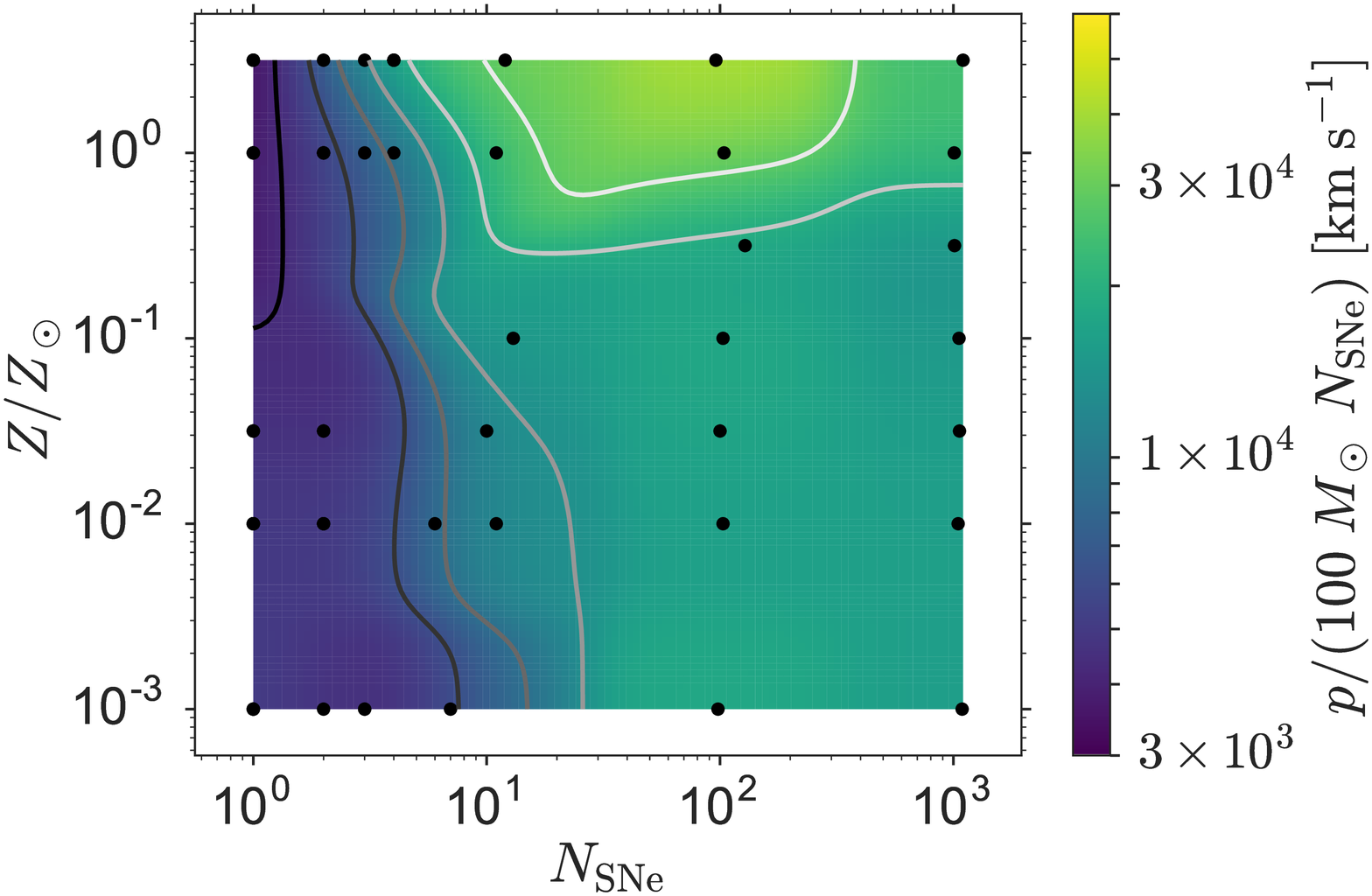}
\caption{
The same as the \autoref{fig:momentum:contour:density}, except now allowing metallicity to vary while holding density fixed at 1.33 $m_\mathrm{H}$ cm$^{-3}$. The top contour level shown in \autoref{fig:momentum:contour:density} is not shown here, as the dynamic range of the data is not as large.
}
\label{fig:momentum:contour:metallicity}
\end{figure}

\begin{figure}
\includegraphics[width=\columnwidth]{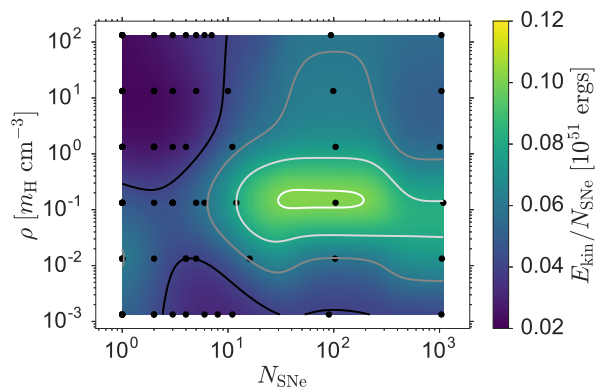}
\caption{
Same as \autoref{fig:momentum:contour:density}, except now the color image shows final kinetic energy with 4 greyscale contours linearly spaced between $2 \times 10^{49}$ and $1.2 \times 10^{50}$ ergs (exclusive). 
}
\label{fig:energy:contour:density}
\end{figure}

\begin{figure}
\includegraphics[width=\columnwidth]{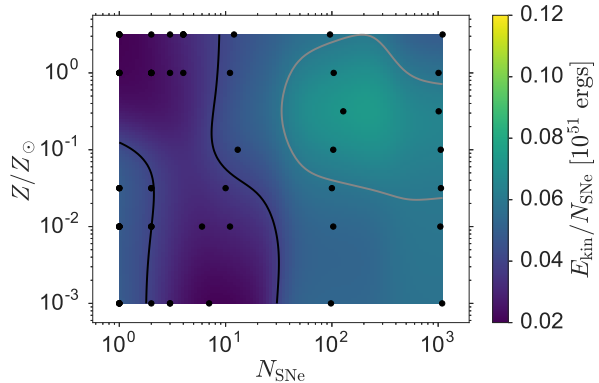}
\caption{
The same as \autoref{fig:energy:contour:density}, except now allowing metallicity to vary while holding density fixed at 1.33 $m_\mathrm{H}$ cm$^{-3}$. The top two contour levels shown in \autoref{fig:energy:contour:density} are not shown here, as the dynamic range of the data is not as large.
}
\label{fig:energy:contour:metallicity}
\end{figure}

As an example of how these results vary across our parameter space, we plot the asymptotic momentum per SN in two cuts through this parameter space in \autoref{fig:momentum:contour:density} (momentum per SN as a function of $\rho$ and $N_\mathrm{SNe}$ at fixed $Z$) and \autoref{fig:momentum:contour:metallicity} (momentum per SN as a function of $Z$ and $N_\mathrm{SNe}$ at fixed $\rho$).  In \autoref{fig:energy:contour:density} and \autoref{fig:energy:contour:metallicity} we provide analogous figures for the asymptotic kinetic energy, which is typically 1--10\% of the injected SN energy.

\section{The Momentum Budget of Clustered Supernovae}
\label{section:momentum}

\autoref{fig:momentum:contour:density} and \autoref{fig:momentum:contour:metallicity} show significant structure in the momentum as a function of number of SNe, gas metallicity and density.  In particular, we note three behaviors: (1) For fixed density and metallicity, starting with a few SNe the momentum per SN initially increases with increasing number of SNe, reaches a maximum between 10-100 SNe (the exact location depends on density and metallicity), then decreases. (2) For a few SNe, the momentum per SN increases with decreasing density and gas metallicity. (3) For many SNe the opposite is true, as momentum per SN increases with increasing density and, to a smaller extent, metallicity.  In this section we show how these primary behaviors are a consequence of clustered SNR evolution falling into one of two physical regimes: the small-$N$ regime and the superbubble regime.

\subsection{Qualitative Analysis}
\label{section:momentum:qualitative}

\subsubsection{The Small-N Regime}
\label{section:momentum:qualitative:few}

\begin{figure}
\includegraphics[width=\columnwidth]{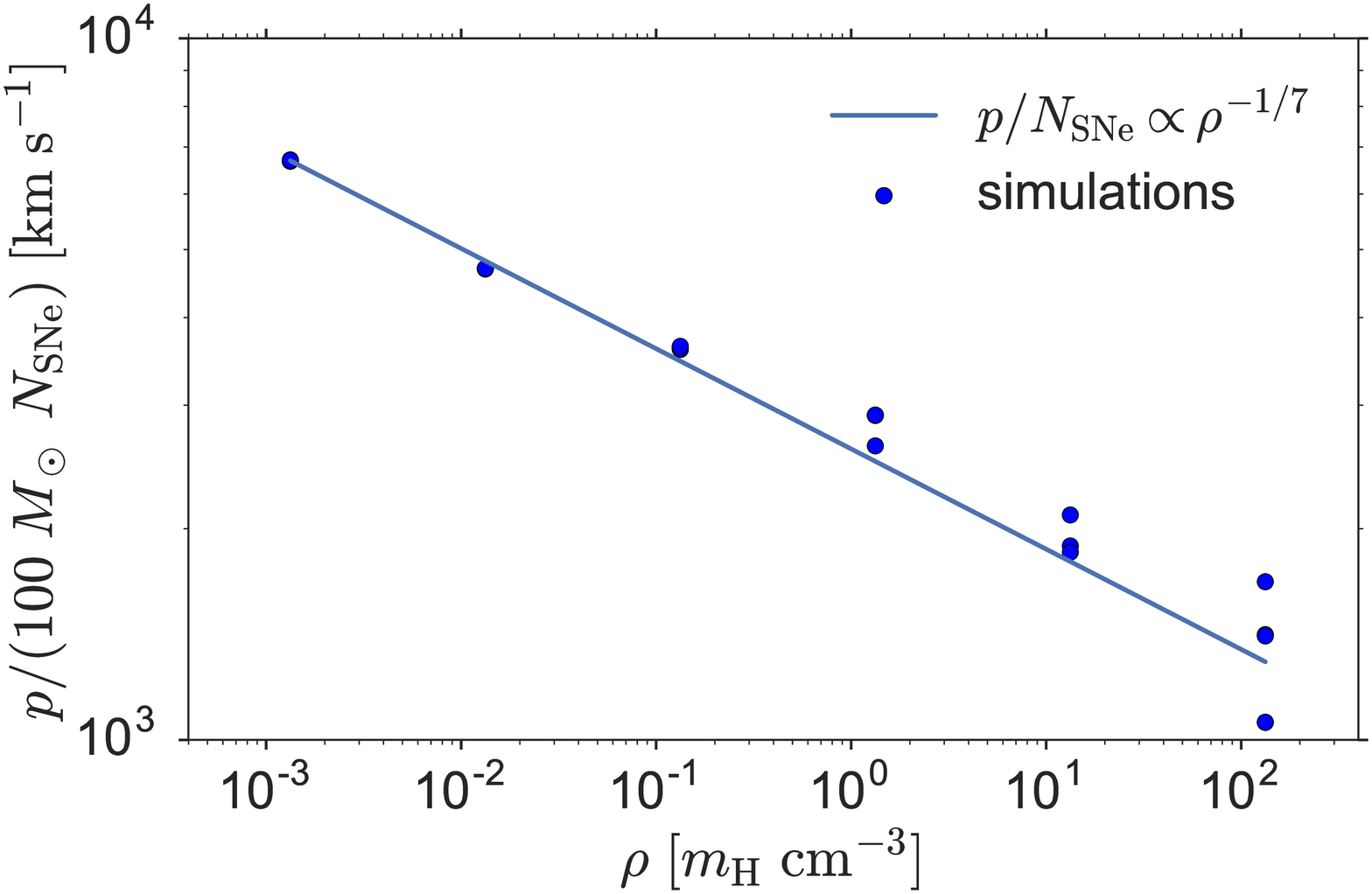}
\caption{
The scaling of momentum with background density, for $Z=Z_\odot$ and $N_\mathrm{SNe}=1$, compared to the $p \propto \rho^{-1/7}$ scaling (normalized to the mean momentum of the lowest density clusters) expected for isolated SNe in a homogeneous background \citep{1988ApJ...334..252C}.
}
\label{fig:momentum:scaling:density:few}
\end{figure}

To understand how SN momentum budgets act when the number of clustered SNe is relatively small, we start with its limiting case: single, isolated SNe. Feedback from isolated SNe has been well explored, as discussed in \autoref{section:intro}. In particular, \citet{1998ApJ...500...95T} found that lower gas metallicities and densities resulted in higher energy feedback, which is what we expect physically; lower gas metallicity and density results in weaker cooling, sapping less energy from a SNR, increasing the amount of energy feedback. This is also expected to apply for momentum feedback, and in \autoref{fig:momentum:scaling:density:few} we show that our results match the scaling between momentum and density expected by \citet{1988ApJ...334..252C}.

\begin{figure}
\includegraphics[width=\columnwidth]{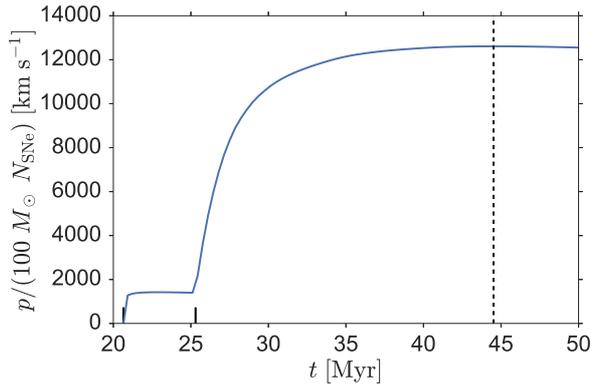}
\caption{
The evolution of the momentum per SN of a $Z = Z_\odot$, $\rho = 1.33 $ $m_\mathrm{H}$ cm$^{-3}$ and $N_\mathrm{SNe} = 2$ cluster. The moment of maximal momentum is marked by the vertical dashed black line; the times of SNe are denoted by solid black ticks. 
}
\label{fig:momentum:evolution:few}
\end{figure}

As the number of SNe increases, the picture is similar to a series of isolated SNe, but with each successive SN occurring in a lower density bubble. As discussed above, this leads to progressively more efficient momentum production as the region is progressively evacuated. \autoref{fig:momentum:evolution:few} illustrates this process directly, by plotting the momentum versus time for a simulation in which two SNe occur. The first SN occurs 20 Myr after cluster formation and its remnant quickly asymptotes to a momentum $\approx 3 \times 10^5$ $M_\odot$ km s$^{-1}$, in agreement with the usual value found for single SNe.  The second SN occurs 5 Myr later, and thanks to the vastly lower density inside the bubble, experiences much smaller radiative losses. This leads it to inject $\approx 2 \times 10^6$ $M_\odot$ km s$^{-1}$ of momentum by the time the momentum peaks, which is almost 10 times more momentum than injected by the first SN.

This breaks down for the clusters with the fewest SNe embedded in the highest density backgrounds, which behave more like multiple, isolated SNe.  As density increases, SNRs evolve more rapidly, quickly cooling, lowering their internal pressure, and then being crushed by the pressure of the surrounding ISM.  For the most dense gas with the fewest SNe, the SNR bubbles can be destroyed between SNe. As subsequent SNe are no longer occurring in the bubble of previous SNe, we believe clustering effects should be minimal.

Unfortunately, since the bubble has collapsed before all SNe have been injected, our numerical methods break down due to the reverse shock propagating all the way to the origin and undergoing unphysical reflection (see \autoref{section:results}). We are therefore forced to exclude this regime from our analysis. In our three-dimensional parameter space, the excluded region is roughly defined by the parameters
\begin{align*}
	\rho \geq 1.33 \; m_\mathrm{H} \text{ cm}^{-3} && \mathrm{and} && N_\mathrm{SNe} < 10 && \mathrm{and} && Z < 0.1 Z_\odot.
\end{align*}
While we cannot simulate this part of parameter space directly, the above analysis suggests that there should be no clustering effects present in it, and thus for the purpose of subgrid modeling it is likely safe to adopt a momentum budget of $3 \times 10^5$ $M_\odot$ km s$^{-1}$ per SN, the same as in the isolated SN regime.

\subsubsection{The Superbubble Regime}
\label{section:momentum:qualitative:many}

While the few SNe model predicts that momentum efficiency increases as the number of SNe increases, our data show a turnover after about 10-100 SNe, beyond which the momentum efficiency begins to drop as the number of SNe increases.  This can be understood within the framework of a superbubble powered by a continuous wind (see \autoref{fig:momentum:evolution:many} for an example of the momentum evolution of a large cluster). As more SNe occur, the bubble density decreases while the bubble temperature increases, both of which lead to less efficient cooling. While this leads to strong momentum feedback for a few SNe, it eventually saturates; if most of the energy is already being retained, suppressing cooling even further will only have a marginal effect.

\begin{figure}
\includegraphics[width=\columnwidth]{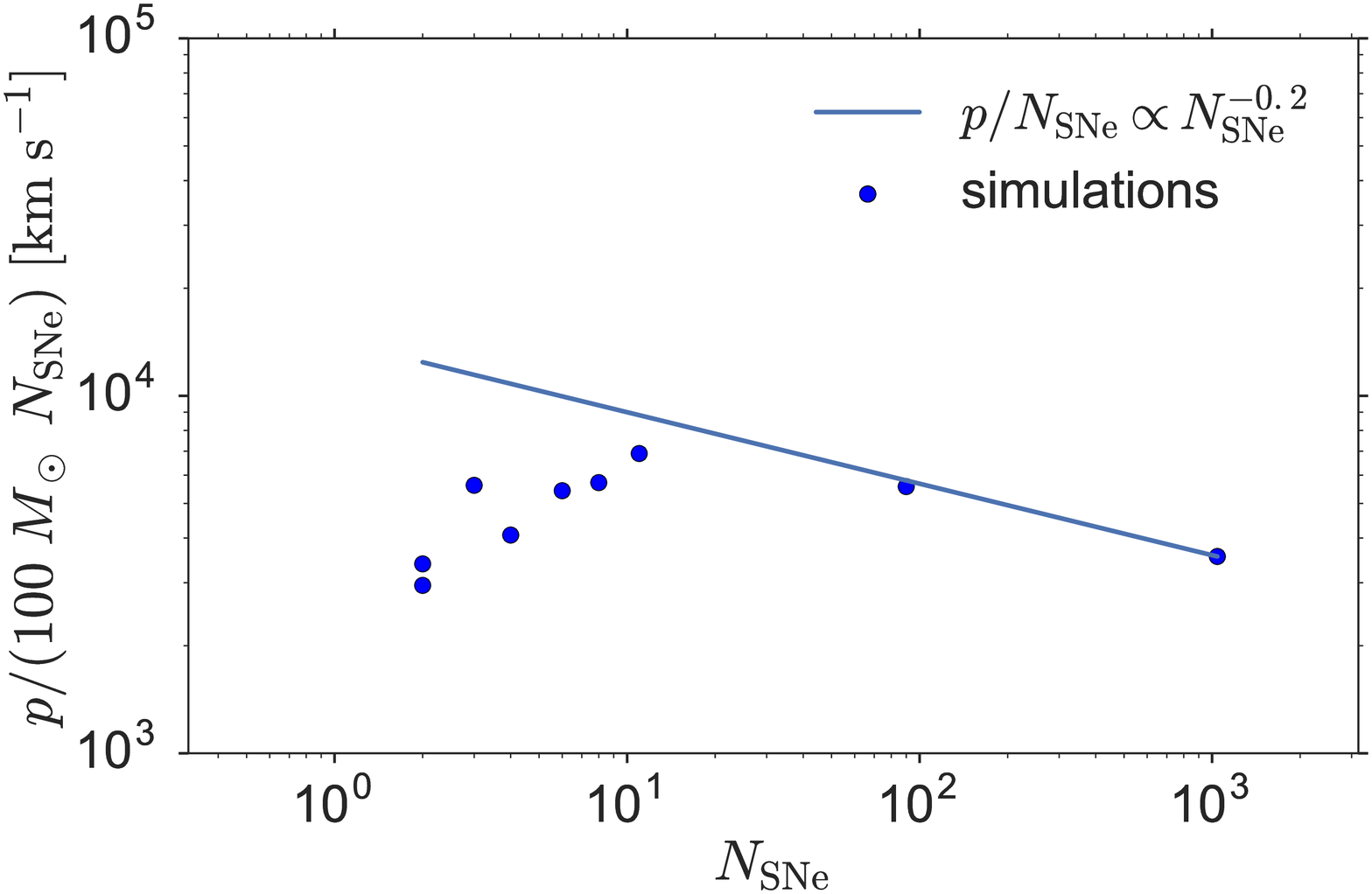}
\caption{
The scaling of momentum per SN with number of SNe, evaluated at the time of the last SN for each cluster. The clusters shown all have solar metallicity and the lowest density simulated, $1.33 \times 10^{-3}$ $m_\mathrm{H}$ cm$^{-3}$. We plot the theoretical scaling for an adiabatic superbubble (\autoref{eq:scaling:theoretical:lastSN}), normalized to the cluster with the most SNe (the cluster which is expected to best correspond to the adiabatic case).
}
\label{fig:momentum:scaling:last_SN}
\end{figure}

\citet{1975ApJ...200L.107C} provide a simplified bubble model which allows us to begin to understand superbubble evolution.  They assume a constant energy injection rate, but if that energy is injected over a period of time that is the same for all clusters (effectively assuming stellar evolution models do not depend strongly on metallicity or density), then their approach is also valid for an energy injection rate that is a power law with respect to time. Using their bubble solution, we can find the momentum per SN at the time of the last SN:
\begin{equation}
	p(t_\text{last SN})/N_\mathrm{SNe} \propto N_\mathrm{SNe}^{-0.2} \rho_0^{0.2} .
	\label{eq:scaling:theoretical:lastSN}
\end{equation}
We compare this predicted scaling to our numerical results for the lowest density simulated (thereby ensuring we are as close as possible to the adiabatic limit) in \autoref{fig:momentum:scaling:last_SN}. As the plot shows, the analytic scaling is in reasonable agreement with the numeric results.

\begin{figure}
\includegraphics[width=\columnwidth]{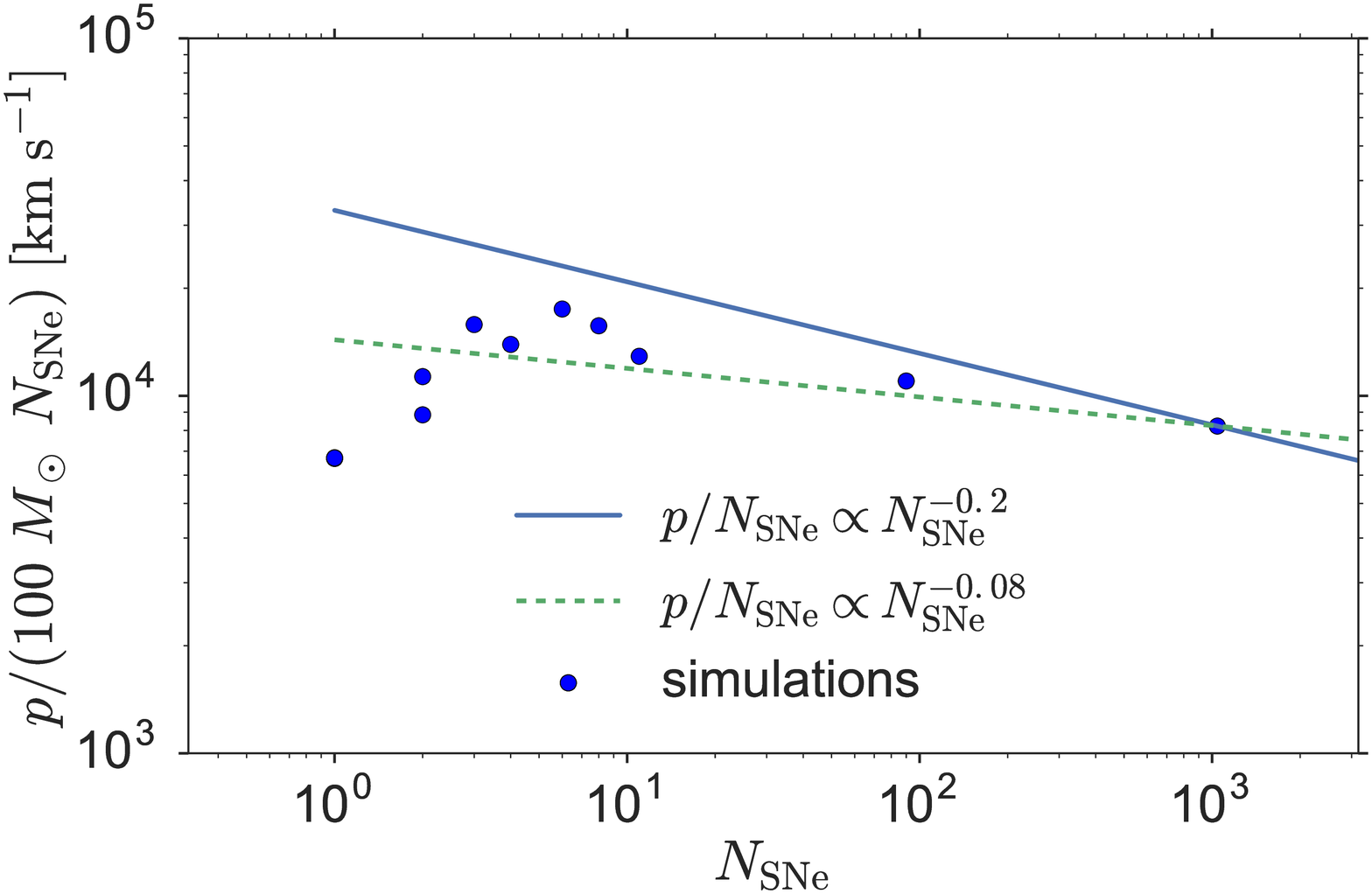}
\caption{
The scaling of asymptotic momentum per SN with number of SNe. These are the same clusters as those shown in \autoref{fig:momentum:scaling:last_SN} ($1.33 \times 10^{-3}$ $m_\mathrm{H}$ cm$^{-3}$; $Z = Z_\odot$) but evaluated at a different time.
We plot theoretical scalings for an adiabatic superbubble at time of the last SN (blue solid line; \autoref{eq:scaling:theoretical:lastSN}) and at the time the interior pressure equals the exterior pressure (green dashed line; \autoref{eq:scaling:theoretical:final}).
}
\label{fig:momentum:scaling:converged}
\end{figure}

\begin{figure}
\includegraphics[width=\columnwidth]{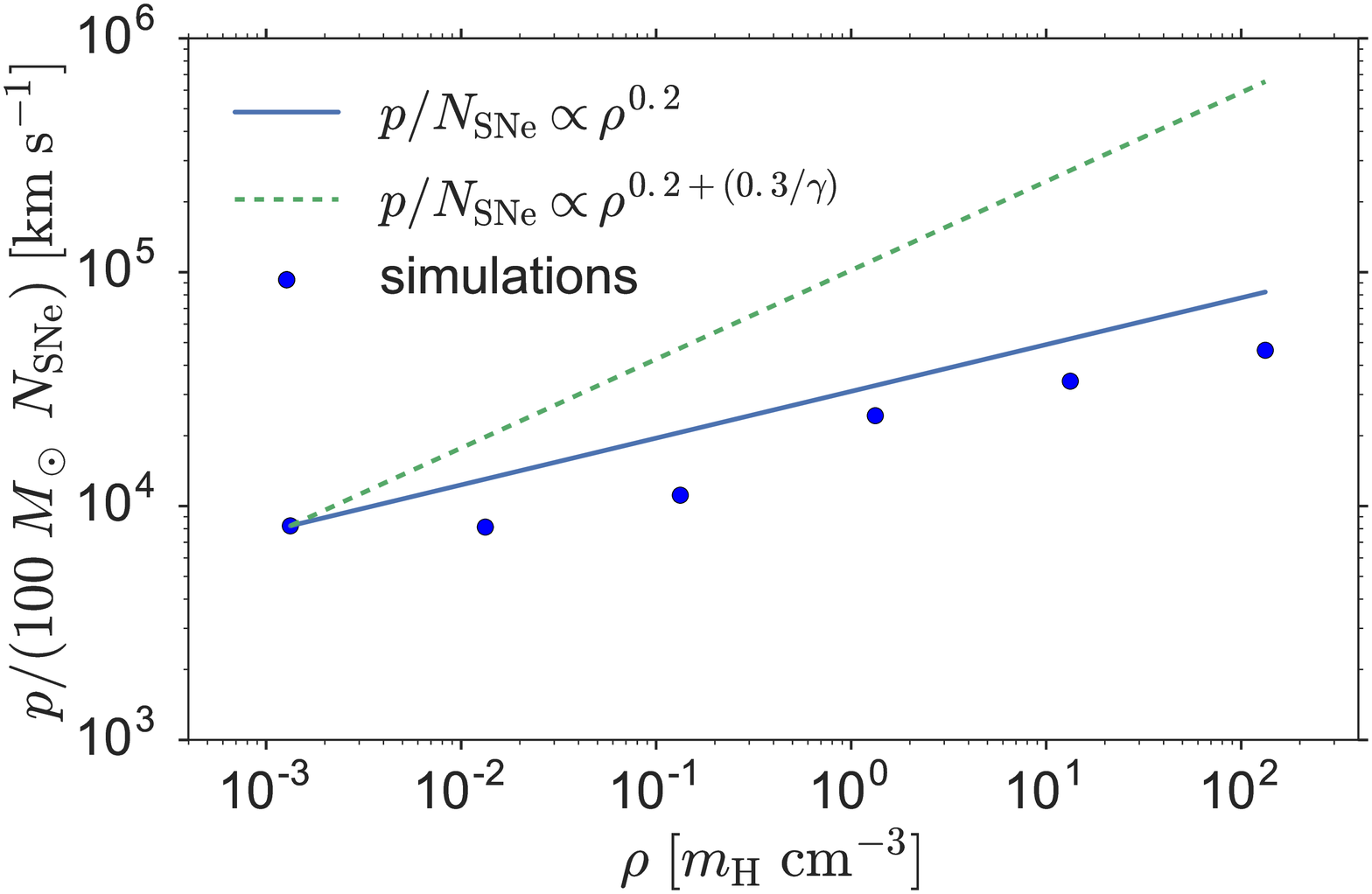}
\caption{
The scaling of asymptotic momentum per SN with background density, for $Z=Z_\odot$ and $M_\mathrm{cluster} = 10^5$ $M_\odot$ ($N_\mathrm{SNe} \approx 10^3$) clusters, compared to the superbubble predictions for the end of the SNe injection phase ($p/N_\mathrm{SNe} \propto \rho^{0.2}$) and when the interior pressure equals the exterior pressure ($p/N_\mathrm{SNe} \propto \rho^{0.2 + (0.3/\gamma)}$).
}
\label{fig:momentum:scaling:density:many}
\end{figure}

As shown in \autoref{fig:momentum:evolution:many}, a significant amount of momentum evolution occurs after the last SN. During this phase, the superbubble expands adiabatically until the bubble pressure equals the ISM pressure, $P_0$, at which point the shell's momentum reaches a maximum, since the pressure gradient switches direction. For an adiabatic index $\gamma=5/3$ for the gas inside the SNR, this results in a final momentum per SN
\begin{align}
	p_\mathrm{final}/N_\mathrm{SNe} &\propto P_0^{-1/(2 \gamma)} N_\mathrm{SNe}^{-0.2 + (0.2/\gamma)} \rho_0^{0.2 + (0.3/\gamma)} \nonumber
	\\
	&\approx P_0^{-0.3} N_\mathrm{SNe}^{-0.08} \rho_0^{0.38}
	\label{eq:scaling:theoretical:final}
\end{align}
This analytic scaling with respect to number of SNe can be compared to our numeric data in \autoref{fig:momentum:scaling:converged}; the scaling with respect to gas density is shown in \autoref{fig:momentum:scaling:density:many}.

It is not surprising that these scalings are not perfect;  \citet{2014MNRAS.443.3463S} predict that even $10^3$ SNe are not enough to satisfy assumptions behind models like those of \citet{1975ApJ...200L.107C}. Specifically, \citet{2014MNRAS.443.3463S} predict no wind-dominated region (where $\rho \propto r^{-2}$) which ends in a stable termination shock before the pressure-dominated bubble begins; these predictions are in agreement with our results (see \autoref{fig:structure:many:lastSN} and \autoref{fig:structure:many:peakmomentum}).  Our simulations do not satisfy all of the assumptions of superbubble models like those of \citet{1975ApJ...200L.107C}; these models are sufficient for a qualitative analysis, but they are insufficient for a quantitative understanding.

\subsection{Quantitative Model}
\label{section:momentum:model}

Informed by the qualitative understanding developed in \autoref{section:momentum:qualitative}, we now construct a quantitative parametric model which we constrain using our simulation results (\autoref{tab:results}).  In both the small-$N$ and superbubble limits, we expect the results to behave like a power law in number of SNe, gas density and metallicity, but we expect these to be different power laws.  Therefore, we choose to construct a model of two power laws with a smooth break. In the few SNe (small-$N$) limit, we use a model of the form
	\begin{multline}
	\left(\frac{p}{N_\mathrm{SNe}}\right)_\mathrm{few} = \left( \frac{p}{N_\mathrm{SNe}} \right)_{0,\mathrm{few}} \left(\frac{Z}{Z_\odot}\right)^{\eta_{Z,\mathrm{few}}} \\  \times \left(\frac{\rho}{m_\mathrm{H} \; \mathrm{cm}^{-3}}\right)^{\eta_{\rho,\mathrm{few}}} \left(\frac{N_\mathrm{SNe}}{1}\right)^{\eta_{N,\mathrm{few}}}
	\end{multline}
and in the many SNe (superbubble) limit we use a similar form,
	\begin{multline}
	\left(\frac{p}{N_\mathrm{SNe}}\right)_\mathrm{many} = \left( \frac{p}{N_\mathrm{SNe}} \right)_{0,\mathrm{many}} \left(\frac{Z}{Z_\odot}\right)^{\eta_{Z,\mathrm{many}}} \\  \times \left(\frac{\rho}{m_\mathrm{H} \; \mathrm{cm}^{-3}}\right)^{\eta_{\rho,\mathrm{many}}} \left(\frac{N_\mathrm{SNe}}{1000}\right)^{\eta_{N,\mathrm{many}}}
	\end{multline}
which are smoothly combined using:
\begin{align}
	\frac{p}{N_\mathrm{SNe}} &= \frac{\left(\frac{p}{N_\mathrm{SNe}}\right)_\mathrm{few} \left(\frac{p}{N_\mathrm{SNe}}\right)_\mathrm{many}}{\left(\frac{p}{N_\mathrm{SNe}}\right)_\mathrm{few}+\left(\frac{p}{N_\mathrm{SNe}}\right)_\mathrm{many}} \label{eq:best_fit}
	\\
	&\approx \mathrm{min}\left[ \left(\frac{p}{N_\mathrm{SNe}}\right)_\mathrm{few},\left(\frac{p}{N_\mathrm{SNe}}\right)_\mathrm{many} \right]
\end{align}

Assuming our simulation results have a random additive gaussian noise of variance $\sigma^2$, we can construct a gaussian likelihood function for the results of each simulation. Even though $\sigma^2$ is unknown, this allows us to determine a maximum likelihood estimate (MLE) for our best-fitting model parameters. We would also like to understand the uncertainties in those parameters.  For that we need the posterior, which through Bayes' theorem requires a prior distribution, $\pi$ on those parameters. Since we do not have strong prior information on most of these parameters, we choose uniform, independent priors on our parameters: $\log \left( \sigma^2 \right)$, $\log\left( p/N_\mathrm{SNe} \right)_{0,\mathrm{few}}$, $\eta_{Z,\mathrm{few}}$, $\eta_{\rho,\mathrm{few}}$, $\eta_{N,\mathrm{few}}$,  $\log\left( p/N_\mathrm{SNe} \right)_{0,\mathrm{many}}$, $\eta_{Z,\mathrm{many}}$, $\eta_{\rho,\mathrm{many}}$, $\eta_{N,\mathrm{many}}$.

Combining this prior with a gaussian likelihood for our data results in the posterior distribution of our model parameters.  We sample this posterior distribution using a Markov Chain Monte Carlo (MCMC) scheme, using Gibbs sampling to draw samples of $\sigma^2$ from an inverse gamma distribution and using a Metropolis-Hastings random walk for the remaining parameters.  We use the MLE as the starting guess, discard the first 10000 steps as burn-in steps and save the next 100000 steps. Using these samples, we can now estimate uncertainties on our model parameters: for each parameter we use the median as our best-fitting value, and the 16$^\mathrm{th}$ and 84$^\mathrm{th}$ percentiles as our uncertainty interval (effectively marginalizing over all other parameters) resulting in

\input{bayesian_fit_intervals.tex}

\begin{figure}
\includegraphics[width=\columnwidth]{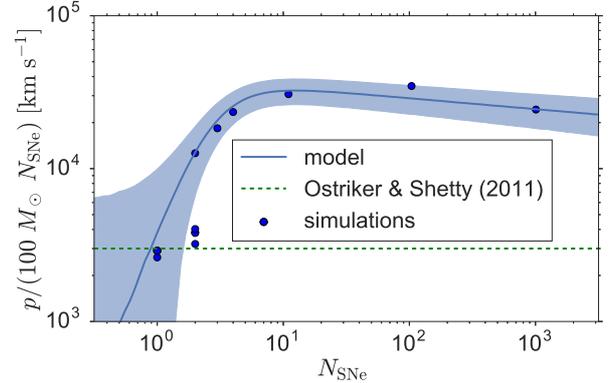}
\caption{
Comparison of a slice of our simulation results ($Z=Z_\odot$, $\rho = 1.33$ $ m_\mathrm{H}$ cm$^{-3}$) to our model with an uncertainty envelop bounding the 16$^\mathrm{th}$ and 84$^\mathrm{th}$ percentiles of our predictive momentum model.  Some slices fit better and some slices fit worse, but overall this is a representative slice. For comparison, we also plot a typical unclustered model, $p/(100 M_\odot N_\mathrm{SNe})= 3000$ km s$^{-1}$ \citep[green dashed line]{2011ApJ...731...41O}.
}
\label{fig:momentum:fit:sample}
\end{figure}

Our posterior samples are also useful for estimating the uncertainty in the predicted momentum from a particular cluster.  For a given gas metallicity, density and number of SNe, each posterior sample predicts a slightly different momentum and an uncertainty $\sigma$ on that momentum. For each posterior sample, we then generate realizations of the noise with variance $\sigma^2$. For $N$ posterior samples and $M$ noise realizations, this gives us $N \times M$ samples of the momentum predictive distribution, which allows us to estimate the median and the 16$^\mathrm{th}$ and 84$^\mathrm{th}$ percentiles of the predictive distribution for a given cluster.  We compare this predictive model and its uncertainties to our a subset of our numeric data in \autoref{fig:momentum:fit:sample}.

\section{Discussion}
\label{section:discussion}

Using our numeric results and quantitative model, we can now comment on the significance of these results in the context of previous works.  We first examine the implications of our results for models of momentum-regulated star and galaxy formation in \autoref{section:discussion:implications}. We then compare our results to those of previous authors in \autoref{section:discussion:comparison}, and in \autoref{section:discussion:additional} we discuss the potential importance of physical processes we have omitted.

\subsection{Implications of High Momentum Efficiency of Clustered SNe}
\label{section:discussion:implications}

In models of momentum-driven feedback, the key parameter is $ p/m_* $, the amount of momentum injected per unit mass of stars formed in a given system.  Non-clustered models of SNe 	momentum production usually assume $ p/m_* \approx p/(100 M_\odot N_\mathrm{SNe}) \approx 1000 - 3000$ km s$^{-1}$ (with a weak dependence on density) for a mass $m_*$ of stars formed \citep{2005ApJ...630..167T,2011ApJ...731...41O,2012ApJ...754....2S,2013MNRAS.432..455D,2013MNRAS.433.1970F,2014MNRAS.445..581H,2015ApJ...802...99K,kimm15a,2015arXiv151005650H}. 
For a star cluster with a single SN ($M_\mathrm{cluster} \approx 100 M_\odot$), our best fit model is a little higher than but still consistent with $1000 - 3000$ km s$^{-1}$ given the uncertainties in our model.  For higher mass clusters, the discrepancy becomes significant.  The most extreme difference is found for $M_\mathrm{cluster} = 10^3 - 10^4 M_\odot$ ($N_\mathrm{SNe} = 10^1 - 10^2$), for which our value of $p/m_*$ can be greater than the unclustered value by an order of magnitude (see \autoref{fig:momentum:contour:density} and  \autoref{fig:momentum:fit:sample}).  But these are just the extremes; for a typical distribution of cluster masses found in a galaxy, what is the average effect?

To evaluate the mean value of $ p/m_* $ for star formation on galactic scales, we must integrate our model for individual clusters, $p/M_\mathrm{cluster}$, over a cluster mass function $dN/dM_\mathrm{cluster}$. The resulting mean momentum yield per unit mass of stars formed is
\begin{equation}
\frac{p}{m_*} = \int \left(\frac{p}{M_\mathrm{cluster}}\right) \frac{dN}{d \ln M_{\rm cluster}} \, dM_{\rm cluster}.
\end{equation}
If we adopt a typical mass distribution $dN/dM_{\rm cluster} \propto M_{\rm cluster}^{-2}$ over the range $M_{\rm cluster} = 10^2 - 10^5$ $M_\odot$, comparable to what is observed in nearby galaxies \citep[and references therein]{krumholz14c}, and use our fitting formula (\autoref{eq:best_fit}) to evaluate $p/N_{\rm SNe}$ as a function of $M_{\rm cluster}$ (for $N_{\rm SNe}\approx M_{\rm cluster}/100M_\odot$), this yields a value of $ p/m_* \approx 1-2\times 10^4$ km s$^{-1}$ over the metallicity range $Z/Z_\odot=0.01-1$ and density range $\rho/m_{\rm H} = 0.1-10^5$. This is $\sim 0.5-1$ dex higher than the value usually adopted based on single SN models. This result is only logarithmically sensitive to the adopted limits on the cluster mass function.

This increased momentum yield will significantly alter the conclusions of analytic models in which star formation is regulated primarily by SN momentum input \citep[e.g.,][]{2011ApJ...731...41O, 2012ApJ...754....2S, 2013MNRAS.433.1970F, 2015arXiv151005650H}. The same is true for models where SN momentum is primarily responsible for launching galactic winds \citep[e.g.,][]{2013MNRAS.432..455D,2015arXiv151005650H,2016MNRAS.455..334T}. In general, the higher momentum yield we obtain will shift such models to predict lower star formation rates for fixed galactic surface densities, which may require re-tuning of other parameters to bring the models back into agreement with observed relationships between star formation rate and gas content.\footnote{The situation is more complex for models that include regulation of star formation by FUV radiation instead of or in addition to SN feedback \citep[e.g.,][]{2009ApJ...699..850K,2010ApJ...721..975O,2013MNRAS.436.2747K}. In these models, the effects of enhanced momentum injection will be more modest or absent, depending on the details of the individual model.} The models will also predict stronger outflows, though these are significantly less constrained by observations.

The momentum yield per SN is also a critical input to numerical methods that handle subgrid feedback through explicit momentum injection \citep[e.g.,][]{2011ApJ...743...25K,2014MNRAS.445..581H,kimm15a,2016ApJ...827...28G}. These models should be also be rerun using our updated estimates of the SN momentum yield. Even models that do not use explicit momentum injection, but that attempt to include SN feedback by explicitly resolving the Sedov phase \citep[e.g.,][]{2011MNRAS.417..950H}, may need to be reconsidered, at least for simulations of galaxies large enough for there to be significant numbers of clustered SNe.

\subsection{Comparison to Previous Work and Convergence Study}
\label{section:discussion:comparison}

We find that clustered SNe generally lead to an increase in the momentum injected by SNe, in some cases by an order of magnitude.  The results of previous authors diverge strongly, as noted in \autoref{section:intro}, with some finding that clustering leads to an enhancement in momentum per SN and others finding a decrease, but none finding an increase as large as an order of magnitude. To understand this discrepancy, we need to understand the role of mixing and how it enters various simulations.

In SNe-driven bubbles, the cooling rate plays a significant role in setting the dynamics of the system.  This cooling rate is itself affected by the mixing rate at interfaces between hot diffuse gas (which dominates the thermal energy) and cold dense gas (which is most radiatively efficient).  If the mixing of energy and matter increases, the cooling rate can increase and the final momentum can decrease. This mixing can be increased by both non-physical sources (i.e., due to numerical diffusion) and physical sources (i.e., conduction or hydrodynamic instabilities that transport energy across the contact discontinuity). While these two channels have very different causes, they can have similar effects on the momentum and energy evolution of the system \citep{2016MNRAS.456..710F}.  We will look at these two channels in turn.

\begin{figure}
\includegraphics[width=\columnwidth]{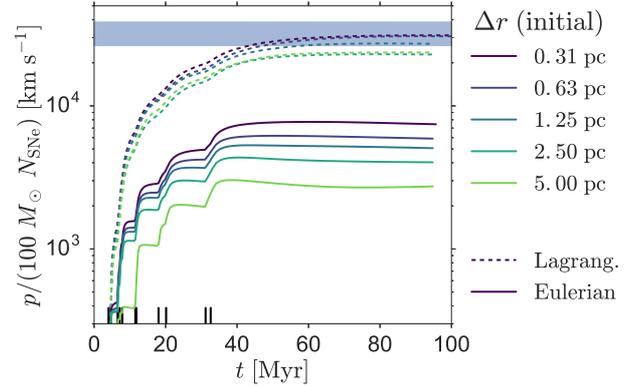}
\caption{
The momentum evolution of a $\rho = 1.33$ $m_\mathrm{H}$ cm$^{-3}$, $Z = Z_\odot$, $M_\mathrm{cluster} = 10^3$ $M_\odot$ ($N_\mathrm{SNe} = 11$) cluster, rerun with a range of initial resolutions, using both Eulerian and Lagrangian methods. The asymptotic momentum predicted by our model is shown by the blue horizontal band which bounds the 16$^\mathrm{th}$ and 84$^\mathrm{th}$ percentiles of the predictive distribution.
}
\label{fig:momentum:evolution:10:resolution}
\end{figure}

\begin{figure}
\includegraphics[width=\columnwidth]{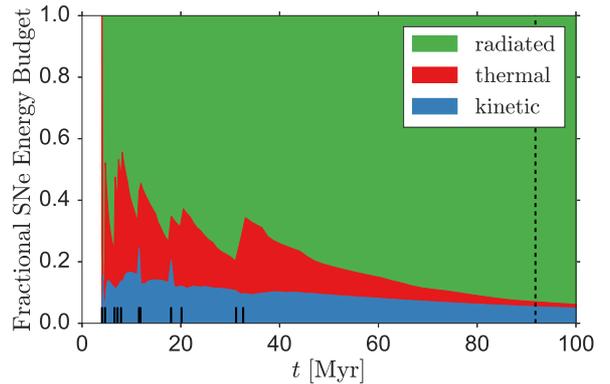}
\caption{
Same as \autoref{fig:energy:evolution:1000}, except now for a $\rho = 1.33$ $m_\mathrm{H}$ cm$^{-3}$, $Z = Z_\odot$, $M_\mathrm{cluster} = 10^3$ $M_\odot$ ($N_\mathrm{SNe} = 11$) cluster, evolved using Lagrangian methods with an initial resolution of $0.6$ pc.
}
\label{fig:energy:evolution:10}
\end{figure}

\begin{figure}
\includegraphics[width=\columnwidth]{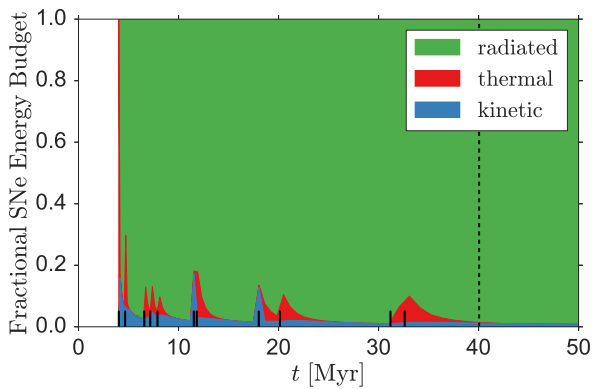}
\caption{
Same as \autoref{fig:energy:evolution:10}, except with the resolution degraded to 2.5 pc, and using a fixed Eulerian mesh.
}
\label{fig:energy:evolution:10:lowres}
\end{figure}

Hydrodynamic solvers that advect mass between adjacent cells fundamentally introduce mixing errors.  These errors can be decreased by improving the resolution or decreasing the mass advected between cells (as we have done by moving our numerical mesh with the fluid).  In order to understand how the choice of resolution and numerical methods affects our results, we re-ran one of our clusters ($\rho=1.33$ $m_\mathrm{H}$ cm$^{-3}$, $Z=Z_\odot$, $M_\mathrm{cluster} = 10^3 M_\odot$, $N_\mathrm{SNe} = 11$; this particular cluster will be useful in later comparisons to \citet{2015ApJ...802...99K}) at a number of initial resolutions, using both a fixed mesh and a moving mesh.  (A fixed mesh corresponds to Eulerian hydrodynamics -- fixing $w=0$ in our methods discussed in \autoref{section:methods} -- which is less accurate and more diffusive than our pseudo-Lagrangian methods.)  We show the results in \autoref{fig:momentum:evolution:10:resolution} and note a few key observations. 
First, the Lagrangian runs appear to be converged (within the uncertainties of our predictive model) by the resolution used for this cluster in our parameter study (an initial resolution of $0.6$ pc).
Second, the Eulerian runs introduce larger errors (as expected) and converge much more slowly.  This suggests that Eulerian and low resolution simulations could have greater errors than high resolution, Lagrangian simulations.
We can better understand these errors by comparing the energy evolution of a high resolution Lagrangian run (\autoref{fig:energy:evolution:10}) and a low resolution Eulerian run (\autoref{fig:energy:evolution:10:lowres}). While the same amount of energy is injected for each SN in both simulations, that energy is radiated away much more rapidly in the low resolution (Eulerian) simulation, draining the bubble of the energy which drives the momentum growth seen in \autoref{fig:momentum:evolution:10:resolution}. (This connection between cooling time and resolution for multiple SNe was also found by \citet{2013A&A...550A..49K}.) The amount of mixing can significantly impact the final momentum, but given the convergence seen in \autoref{fig:momentum:evolution:10:resolution}, the results we have obtained appear to be converged.

While our resolution study suggests that high resolution, low diffusion simulations are required to achieve accurate results, that conclusion might not apply if there are stronger, physical diffusive processes are present \citep{2016MNRAS.456..710F}. For a SNR or a superbubble, a number of processes can mix gas between the hot bubble interior and the cool shell; for a review see Appendix B of \citet{2016MNRAS.456..710F}.
Our 1D code cannot simulate many of these processes directly\footnote{There exist prescriptions for approximating mixing instabilities in 1D codes \citep[e.g.][]{2016ApJ...821...76D}, but none were incorporated in our work.}, but higher dimensional simulations can.  In order to test the effects of these more complex mixing interfaces, we will compare our results to existing 3D simulations of multiple SNe in an inhomogeneous background.

\citet{2015MNRAS.450..504M}, \citet{2015MNRAS.451.2757W} and \citet{2015ApJ...802...99K} all tested the effects that a turbulent or multi-phase background might have on SNR evolution.  For 1 SN, they all found that an inhomogeneous background makes a relatively small difference: a change of less than 60\%.  They also test multiple SNe in an inhomogeneous background, but none compare the results to multiple SNe in a homogeneous background.  So in order to understand the effect of mixing in the case of multiple SNe, we will compare one of our clusters ($\rho=1.33$ $m_\mathrm{H}$ cm$^{-3}$, $Z=Z_\odot$, $M_\mathrm{cluster} = 10^3 M_\odot$, $N_\mathrm{SNe} = 11$; the cluster used in our resolution study) with a multi-phase multiple SNe cluster from \citet{2015ApJ...802...99K} ($\rho=1.4$ $m_\mathrm{H}$ cm$^{-3}$, $Z=Z_\odot$, $N_\mathrm{SNe} = 10$). Note that although these clusters were chosen to be as similar as possible (except with a difference in background media), they also differ in SN delay time distributions, ejecta and mass loss prescriptions.  Nevertheless, we can compare our results to those of \citet{2015ApJ...802...99K} and find that their cluster cools much more rapidly than ours, leading to an asymptotic momentum per SN which is a factor of 20 lower than ours.  

We believe that this difference is due to physical mixing that is present in their simulations but not ours.  While our results can be brought into agreement with theirs by degrading our resolution and using less accurate numerical methods (as shown by our convergence study), subsequent simulations have shown their results to be converged with respect to resolution (Kim, Ostriker, \& Raileanu 2016, in prep.).  If a two-phase background results in significantly increased physical mixing, that would explain how their converged simulations could appear similar to our less-accurate, unresolved simulations: a strong source of physical mixing can appear similar to strong artificial mixing \citep{2016MNRAS.456..710F}, while being less sensitive to the resolution.  

One might ask at this point whether the enhanced momentum injection we find is solely a result of our use of 1D simulations, which necessarily suppressing mixing. Such a conclusion might be comforting, but is far from warranted. The overall lesson to draw from this comparison is that the cooling rate and momentum budget for bubbles produced by multiple SNe is exquisitely sensitive to the amount of mixing, whether physical or numerical.
For a homogeneous background, very high resolution is required to get a converged value for the asymptotic momentum.  This resolution requirement can be rendered irrelevant if strong, physical mixing occurs, allowing convergence at much lower resolution. But accurate results require more than just convergence; to be confident in the accuracy of a set of results, one must be confident that the physical mixing processes have been properly captured. In their multiple SNe simulations \citet{2015ApJ...802...99K} include a two-phase background, but not magnetic fields, which are known to suppress mixing across contact discontinuities in other contexts \citep[e.g.,][]{Markevitch07a}. Thus their results should probably be regarded as lower limits. \citet{2014MNRAS.443.3463S} do include magnetic fields, but only the context of a uniform medium. Given the state of the field, and the resolution requirements we have obtained in the uniform case, it seems clear that there is an urgent need to re-examine the momentum budget of clustered SNe in a multi-dimensional context, properly including all the mechanisms that can both enhance and suppress mixing.

\subsection{The Effects of Additional Physics}
\label{section:discussion:additional}

In order to render the problem as clean as possible, we have focused only on type II SN feedback in a uniform medium. We now consider how other physical processes that we have heretofore neglected might alter our results.

\subsubsection{Pre-SN Radiative Feedback}
\label{section:discussion:additional:pre-SN}

Before any SNe occur, we expect pre-SN feedback to already be sculpting the region.  In particular, ionizing radiation from young stars can create an overpressured, expanding bubble, lowering the density in which SNe occur. In addition, expansion of the H~\textsc{ii} region will by itself add some momentum to the gas. Neither effect is included in our model, and we would like to understand if this significantly biases our results.

The ionizing luminosity of a cluster of mass $M_\mathrm{cluster}$ is $Q = 10^{49.6}$ $M_\mathrm{cluster} / (10^3 M_\odot)$ s$^{-1}$ \citep{1999ApJS..123....3L}.  These photons will ionize a bubble of gas around the cluster, raising the temperature to $10^4$ K.  For density $\rho < $ $m_\mathrm{H}$ cm$^{-3}$, this ionized region is not much hotter than the background (which has a temperature only slightly below $10^4$ K, appropriate for warm neutral gas). This means the ionized bubble will not be significantly over-pressured compared to the background, so it will not expand significantly.  For clusters in backgrounds of density $\rho < $ $m_\mathrm{H}$ cm$^{-3}$, we therefore do not expect pre-SN radiative feedback to affect our results.

For higher densities, the equilibrium temperature of the gas is well below $10^4$ K, so the $10^4$ K ionized bubble is significantly over-pressured compared to its surroundings. This will allow it to expand, lowering the density in which SNe occur.  For uniform density, and neglecting the small range of parameter space where radiation pressure effects will be significant \citep{2009ApJ...703.1352K}, the H~\textsc{ii} bubble radius $r_\mathrm{II}$ will be governed by the classical \citet{1978JRASC..72..349S} solution\footnote{This solution assumes $t \gg t_{S,0}$, otherwise expansion is expected to be negligible.  For the cluster masses considered, this assumption typically fails for $n < 1$ cm$^{-3}$, but we already expected minimal expansion in such backgrounds.} \citep[chapter 7]{2015arXiv151103457K},
\begin{equation}
r_{\rm II} \approx r_{S,0} \left(\frac{7t}{2\sqrt{3} t_{S,0}}\right)^{4/7},
\end{equation}
where $r_{S,0}$ is the Str\"{o}mgren radius at the start of expansion, given by\footnote{Note that this and the following expressions assume that He is singly ionized.}
\begin{align}
r_{S,0} &= \left[\frac{3 Q \mu^2 m_{\rm H}^2}{4(1.1)\pi \alpha_B \rho^2}\right]^{1/3} \nonumber \\
&= 3.1 Q_{49}^{1/3} n_2^{-2/3} T_{\mathrm{II}, 4}^{0.272 + 0.007 \ln T_{\mathrm{II}, 4}} \mbox{ pc},
\end{align}
and where $Q_{49} = Q / 10^{49}$ s$^{-1}$, $\mu = 1.33$, $n_2 = \rho/(\mu m_{\rm H})/100$ cm$^{-3}$, $T_\mathrm{II}$ is the temperature of the ionized gas, $T_{\mathrm{II}, 4} = T_{\mathrm{II}} / 10^4$ K, $\alpha_B \approx 2.54 \times 10^{-13} T_{\mathrm{II},4}^{-0.8163 - 0.0208 \ln T_{\mathrm{II}, 4}}$ cm$^3$ s$^{-1}$ is the case B recombination coefficient \citep{2011piim.book.....D}, $t_{S,0} = r_{S,0}/c_{\rm II}$,
\begin{equation}
c_{\rm II} = \sqrt{2.2 \frac{k_\mathrm{B} T_{\rm II}}{\mu m_{\rm H}}} = 12 T_{\rm II,4}^{1/2}\mbox{ km s}^{-1},
\end{equation}
is the ionized gas sound speed. The corresponding density inside the H~\textsc{ii} region is
\begin{equation}
\rho_{\rm II} = \rho \left(\frac{r}{r_{S,0}}\right)^{-3/2} = \rho \left(\frac{7t}{2\sqrt{3} t_{\mathrm{S},0}}\right)^{-6/7}.
\end{equation}
The mass and velocity of the swept-up shell are
\begin{eqnarray}
v_{\rm II} & = & \frac{4}{7} \frac{r_{\rm II}}{t} \\
M_{\rm sh} & = & \frac{4}{3} \pi r_{\rm II}^3 \rho.
\end{eqnarray}
We are interested in the properties of the H~\textsc{ii} region at a time of $\approx 4$ Myr, when the first supernova occurs; these are
\begin{align}
r_{\rm II} & = 27 M_{\rm cluster,3}^{1/7} n_2^{-2/7} \left(\frac{t}{4\mbox{ Myr}}\right)^{4/7} T_{\mathrm{II},4}^{0.402 + 0.003 \ln T_{\mathrm{II}, 4}}\mbox{ pc} \\
\frac{\rho_{\rm II}}{\rho} & =  0.077 M_{\rm cluster,3}^{2/7} n_2^{-4/7} \left(\frac{t}{4\mbox{ Myr}}\right)^{-6/7} T_{\mathrm{II},4}^{-0.195 + 0.006\ln T_{\mathrm{II}, 4}}
\label{eq:rhoII} \\
p_{\rm II} & =  7.7\times 10^5 M_{\rm cluster,3}^{4/7} n_2^{-1/7} \left(\frac{t}{4\mbox{ Myr}}\right)^{-3/7} \nonumber \\
& \quad\quad \times T_{\mathrm{II},4}^{1.41 + 0.02 \ln T_{\mathrm{II}, 4}} M_\odot\mbox{ km s}^{-1} \label{eq:pII}
\end{align}
where $M_{\rm cluster,3} = M_{\rm cluster}/10^3M_\odot$, $p_{\rm II}$ is the momentum of the shell. For the rest of this section we assume $T_{\mathrm{II}, 4} = 1$.

Based on these results, for $M_\mathrm{cluster} \approx 100$ $M_\odot$ the extra momentum in the H~\textsc{ii} region ($\approx 2 \times 10^5$ $M_\odot$ km s$^{-1}$) is $\sim 50\%$ smaller than that injected by SNe ($\approx 3 \times 10^5$ $M_\odot$ km s$^{-1}$), and the fractional contribution drops fairly rapidly for more massive clusters thanks to the sublinear scaling of $p_\mathrm{II}$ with $M_\mathrm{cluster}$ (\autoref{eq:pII}). Thus the extra momentum injected directly by the H~\textsc{ii} region is mostly negligible.

The second effect, lowering the density of the medium into which the SNe expand, matters if the region of lowered density encompasses where a SNR would otherwise experience significant cooling: beyond the shell formation radius of the first SN (when the remnant exits the Sedov phase). This radius is \citep[their equation 8, assuming an energy budget of $10^{51}$ ergs per 100 $M_\odot$ of stars]{2015ApJ...802...99K}
\begin{equation}
r_{\rm sh} = 6.4 M_{\rm cluster,3}^{0.29} n_2^{-0.42}\mbox{ pc}.
\end{equation}
Given the weak scaling of both $r_{\rm sh}$ and $r_{\rm II}$ with $M_{\rm cluster}$ and $n$, and the lower coefficient for $r_{\rm sh}$, this means that the presence of an H~\textsc{ii} region can at least potentially affect the evolution at densities $n \gtrsim 1$ cm$^{-3}$.

To quantify the effect of this radiative feedback, we can use the results of \citet{2015MNRAS.451.2757W}, who ran simulations with and without pre-SN ionization for $N_\mathrm{SNe} = 1$ and $\rho \approx 100 $ $m_\mathrm{H}$ cm$^{-3}$.  They found that ionization led to a pre-SNe bubble of density $\rho \approx 10 $ $m_\mathrm{H}$ cm$^{-3}$, which resulted in a final momentum 50\% higher than their simulation without ionization (runs ``HCI'' and ``HC'' respectively.)  This change in momentum can be explained well by re-scaling the results from the non-ionized run to the density within the bubble of the ionized run (using the $p \propto \rho^{-1/7}$ scaling of \citet{1988ApJ...334..252C}), with the added extra momentum injected by the H~\textsc{ii} region directly.  Thus for single SNe, it appears that the relevant density for a momentum feedback model is the ionized bubble density rather than the background density.

Assuming that this conclusion can be extended to the case of multiple SNe, we can quantity the effects of pre-SN ionizing radiation simply by combining the density scaling in \autoref{eq:rhoII} with our best-fitting density dependences, $p\propto \rho^{-0.06}$ in the few-SN regime and $p\propto \rho^{0.14}$ in the superbubble regime. In the few-SN regime, this implies an increase in the momentum yield per SN by a factor of
\begin{equation}
f_{\rm II,few} \approx 1.17 M_{\rm cluster,3}^{-0.017} n_2^{0.034}.
\end{equation}
The corresponding effect in the superbubble regime is a \textit{decrease} in the momentum yield by a factor of
\begin{equation}
f_{\rm II,many} \approx 0.70 M_{\rm cluster,3}^{0.040} n_2^{-0.080}.
\end{equation}

Thus in general we expect that the effect of a pre-SN H~\textsc{ii} region will be to alter the final momentum yield at the tens of percent level, with the sign of the effect depending on the whether we are in the few-SN or the superbubble regime.

\subsubsection{Type Ia SNe}
\label{section:discussion:additional:Ia}

\begin{figure}
\includegraphics[width=\columnwidth]{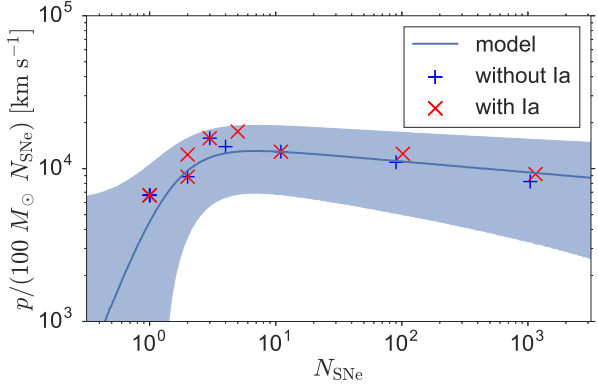}
\caption{
Comparison of simulations with just core-collapse SNe (marked by a blue $+$), and simulations with core-collapse and Type Ia SNe (marked by a red $\times$) for a set of simulations with $Z=Z_\odot$ and $\rho = 1.33 \times 10^{-3}$ $m_\mathrm{H}$ cm$^{-3}$.
}
\label{fig:Ia:lowdensity}
\end{figure}

\begin{figure}
\includegraphics[width=\columnwidth]{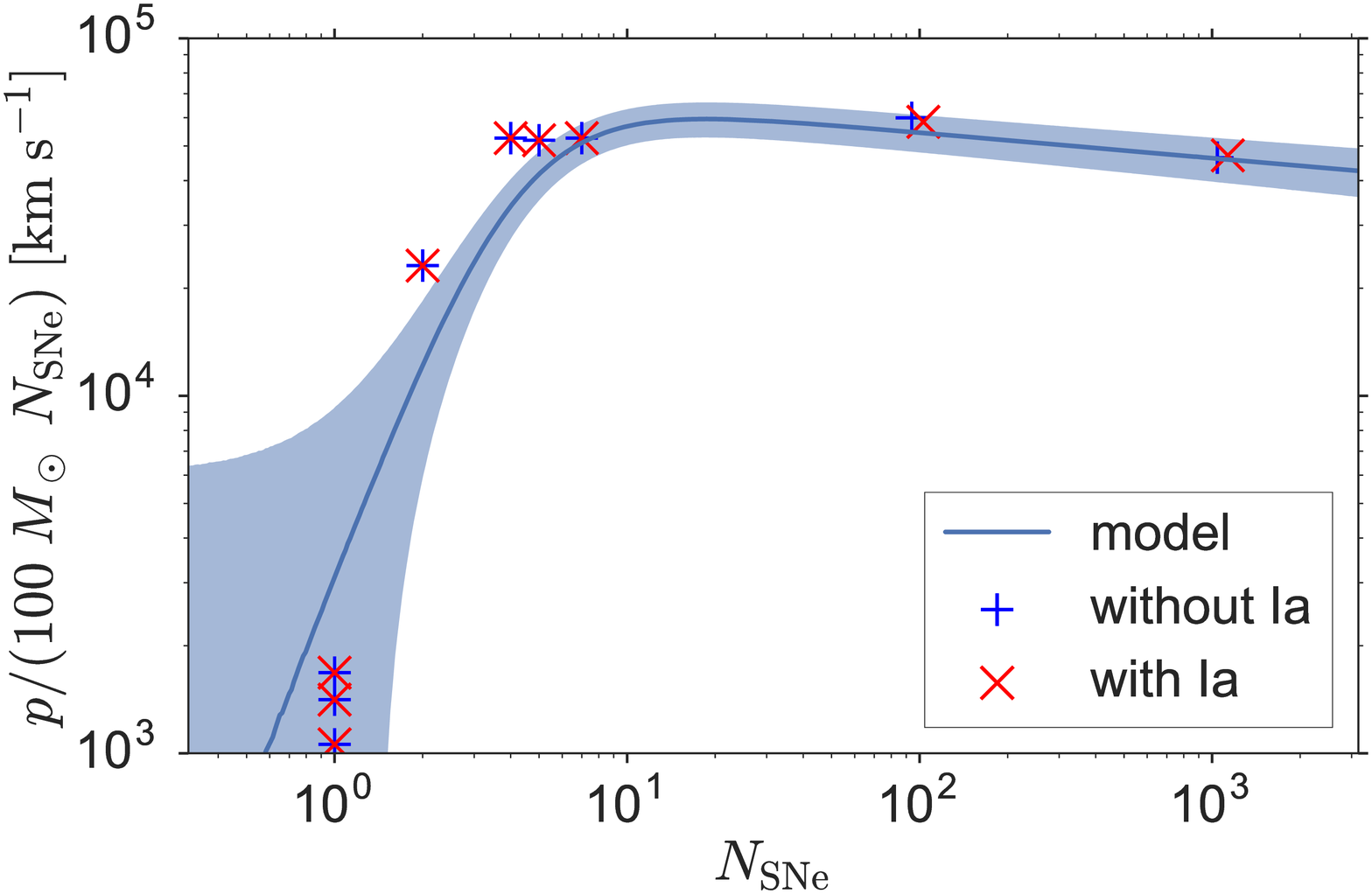}
\caption{
Same as \autoref{fig:Ia:lowdensity}, except now for $\rho = 1.33 \times 10^2$ $m_\mathrm{H}$ cm$^{-3}$ clusters.
}
\label{fig:Ia:highdensity}
\end{figure}

We just discussed pre-SN alterations to our results -- what about changes after the core-collapse SNe occur? In particular, could subsequent Type Ia SNe rejuvenate an old superbubble?

To test this we rerun a subset of our highest and lowest density simulations with Type Ia SNe added after all the core-collapse SNe have occurred.   We add $9.75 \times 10^{-4}$ Type Ia SNe per $M_\odot$ of stars, with the exact number sampled from a Poisson distribution; this rate is taken from \citet{2014ApJS..210...14K}, rescaled to a \citet{Kroupa04012002} IMF.  We draw Type Ia SN delay times  from a $t^{-1}$ distribution, beginning at $t=40$ Myr and extending to 100 Myr.  This is not meant to be a complete accounting of the full effect of Type Ia SNe; most Type Ia SNe occur after much longer delays, and within 40-100 Myr the delay time distribution is poorly constrained \citep{2012MNRAS.426.3282M}; this is simply to test the effects of SNe that might occur while the bubble still exists.

We show results for the low and high density runs in \autoref{fig:Ia:lowdensity} and \autoref{fig:Ia:highdensity}. For relatively short-delay Type Ia SNe, we find that they are consistent with our model if $N_\mathrm{SNe}$ is increased accordingly ($N_\mathrm{SNe} = N_\text{core-collapse} + N_\text{Type Ia}$).  For long-delay SNe we caution against using our model; it is not guaranteed that both the progenitor will remain within the cluster and that the bubble will survive much longer than 100 Myr.

\subsubsection{Self-Gravity}
\label{section:discussion:additional:self-gravity}

Previous studies of SNRs and superbubbles have differed in regards to whether they include or exclude gravitational forces. For example, \citet{2015MNRAS.450..504M} and \citet{2015ApJ...802...99K} do not include gravity, while \citet{1998ApJ...500...95T} and \citet{2015MNRAS.451.2757W} do include self-gravity.  We chose not to include any gravitational forces in our main simulations, and now we estimate what effect that has on our results. In this section we focus on the effects of the self-gravity of the simulated gas, rather than external gravitational forces. 

First, it is useful to understand why we did not include gravity ... in our simulations. This is partly a philosophical choice. The momentum budget we are attempting to compute is frequently used as an input to models of feedback-regulated star formation or wind generation. In such models, the feedback is compared to the force of gravity either analytically or numerically. For this purpose, the momentum budget in which we are interested is that before the effects of gravity are applied; to include gravity would be in effect to double-count it, by inserting it once into the subgrid feedback model and then a second time into the overall model. However, there is also a practical reason that we omit gravity.  Strictly speaking, the self-gravitational potential of an infinite, uniform medium is undefined.  We could assume an external potential dominates, but only if it were spherically symmetric about the cluster, which rules out many potentials of interest, like a galactic potential. Additionally, including a gravitational force, especially self-gravity, would cause our uniform background to collapse.  This was not a problem for \citet{1998ApJ...500...95T}, who simulated much less than a free-fall time, but caused \citet{2015MNRAS.451.2757W} to limit their simulations to 1 Myr in duration. Unfortunately, limiting the duration of our simulations was not an option; we needed to simulate SNe over a period of at least 30 Myr, much longer than a free-fall time for most of our densities. Rather than artificially require a pressure gradient that ensures equilibrium, we chose to exclude gravitational forces.

Still, gravity exists in real systems, so we should try to understand its effect on our work. First, we will use analytic, simplified arguments to predict the effects of self-gravity on our simulations. We then use a simplified prescription to include self-gravity directly in our numeric simulations.  By comparing those results we can begin to understand the effects of self-gravity and the limitations of our analytic model.

For an arbitrarily thin shell dominated by mass swept up from a constant density background the force of self-gravity is
\begin{equation}
	\label{eq:gravity:force:self}
	F_\mathrm{grav} = \frac{G M_\mathrm{shell}^2}{2 R_\mathrm{shell}^2} = \frac{8 \pi^2}{9} G \rho_0^2 R_\mathrm{shell}^4 .
\end{equation}
When this inward force becomes greater than the force exerted by the pressure of the hot bubble, the momentum will stop increasing, ending the simulation (unless the bubble would have already mixed into the ISM and the simulation is already completed). This competing force from the hot bubble gas is
\begin{equation}
	F_\mathrm{gas} \approx 4 \pi R_\mathrm{shell}^2 P_\mathrm{bubble} = 3 (\gamma-1) \frac{E_\mathrm{R,int}}{R_\mathrm{shell}} .
\end{equation}
These two forces become equal at a radius
\begin{align}
	R_\mathrm{max} &= \left( \frac{27}{8 \pi^2} (\gamma-1) \frac{ E_\mathrm{R,int}}{G \rho_0^2} \right)^{1/5} \label{eq:gravity:radius}
	\\
	&\approx  1200 \left(\frac{N_\mathrm{SNe}}{1000} \right)^{1/5}  \left(\frac{\rho_0}{1.33 \times m_\mathrm{H} \text{ cm}^{-3}} \right)^{-2/5} \text{ pc} 
\end{align}
where $R_\mathrm{max}$ is the maximum radius the bubble could reach, having assumed all the injected energy is retained as internal energy ($E_\mathrm{R,int} = N_\mathrm{SNe} \times 10^{51}$ ergs). This provides a simple way to include the effects of gravity via post-processing: using the $R_\mathrm{max}$ determined by \autoref{eq:gravity:radius}, we can truncate a simulation at that radius using the data in \autoref{tab:evolution}.
If the bubble mixes into the ISM at a radius smaller than $R_\mathrm{max}$, then gravity is assumed to have no effect.

\begin{figure}
\includegraphics[width=\columnwidth]{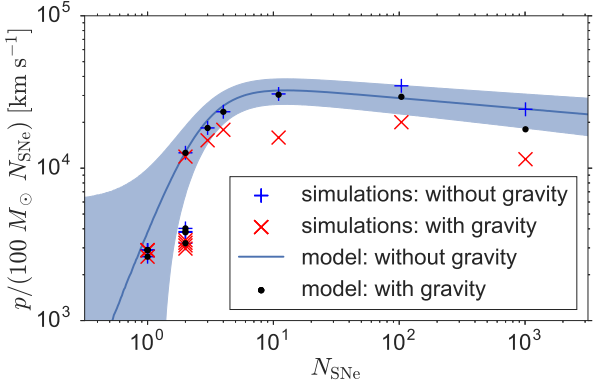}
\caption{
Comparison of simulations with no gravity (marked by a blue $+$) and simulations with gravity applied to the remnant (marked by a red $\times$) for all of our simulations with $Z=Z_\odot$ and $\rho = 1.33 \times 10^{-3}$ $m_\mathrm{H}$ cm$^{-3}$. We also include a simple post-processing prediction for the effect of self-gravity, which truncates our without-gravity simulations if and when they reach $R_\mathrm{max}$ (\autoref{eq:gravity:radius}).
}
\label{fig:gravity:self}
\end{figure}

We can also re-run a subset of simulations including gravity explicitly.  To do this, we calculate the force of self-gravity and the force due to a central point source of mass $M_\mathrm{cluster}$, and use these forces to compute the appropriate source terms for momentum and energy.  These source terms are then only applied to the shocked gas (cells with $r < R_\mathrm{shock}$, where $R_\mathrm{shock}$ is the radius of the overdensity furthest from the center).  By only applying gravity to shocked gas, we are able to approximate our analytic approach (which only considers gravity of the shell), and avoid the problem of our background collapsing.  
This effectively assumes the background is kept in equilibrium by a corresponding thermal or dynamic pressure gradient. We could have explicitly included this pressure gradient in our simulations, but chose not to, so that our with-gravity simulations would better correspond to our without-gravity simulations.  A more sophisticated treatment of gravity using higher dimensional simulations would be worthwhile.

In \autoref{fig:gravity:self} we compare the results for simulations with and without gravity, and the results of our post-processing model which truncates the without-gravity simulations.  We see that including gravity in simulations can decrease the final momentum, and that this decrease can be significant (compared to the uncertainties in our without-gravity model), and as large as a factor of 2.3.  
We also see that the post-processing truncation model typically over-predicts the final momentum (under-predicting the effects of gravity), relative to the simulations which incorporate gravity directly. This is expected; our post-processing model assumes all the SN energy remains as thermal energy, but in our simulations some SN energy is converted into kinetic energy, some into potential energy and most is radiated away.  This causes simulations to stop at smaller radii than predicted, leading to lower final momenta than the post-processing model predicts.

We find that self-gravity can significantly change the final momentum, especially for clusters of many SNe, but that momentum is nevertheless still enhanced by a factor of about 4 compared to the single SN case.  

\subsubsection{Galactic Environment}
\label{section:discussion:additional:external-gravity}

In \autoref{section:discussion:additional:self-gravity} we investigated the effects of self-gravity, but in some cases a galactic gravitational potential might play a larger role in shaping the late-time dynamics of large bubbles.  In this section we will estimate the effects of a galactic gravitational potential, as well as the effects of rotational shear and disc breakout. For each of these effects, we will use a post-processing correction similar to that used in \autoref{section:discussion:additional:self-gravity}: calculate a radius or time where our assumptions break down, and then use the data in \autoref{tab:evolution} to truncate the bubble evolution at that radius or time. 

\begin{figure}
\includegraphics[width=\columnwidth]{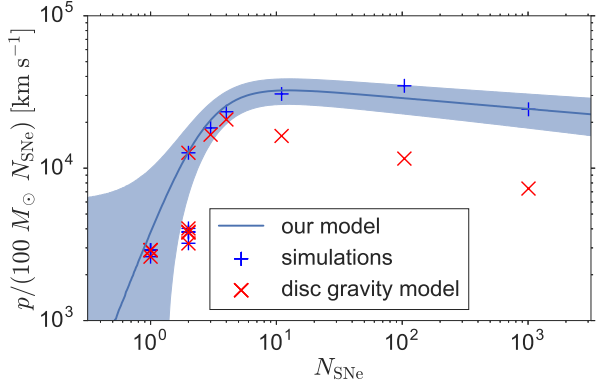}
\caption{
Comparison of simulations with no galactic gravity (marked by a blue $+$) and our simple post-processing model of galactic gravity in a disc for all of our simulations with $Z=Z_\odot$ and $\rho = 1.33 \times 10^{-3}$ $m_\mathrm{H}$ cm$^{-3}$.
}
\label{fig:gravity:disc}
\end{figure}

First, we consider the gravitational force perpendicular to a galactic disc, produced by the galactic gravitational potential. The acceleration due to this force can be written as:
\begin{equation}
	g = \frac{z}{r_\mathrm{g}} \frac{v_\mathrm{c}^2}{r_\mathrm{g}}
\end{equation}
where $v_\mathrm{c}$ is the circular velocity at a galactocentric radius $r_\mathrm{g}$, and $z$ is the distance from the midplane. For simplicity, we will assume all of the mass of the shell is in a plane at height $z = R_\mathrm{shock}$. This results in a force
\begin{equation}
	F_\mathrm{grav} = M_\mathrm{shell} \frac{R_\mathrm{shell}}{r_\mathrm{g}} \frac{v_\mathrm{c}^2}{r_\mathrm{g}}.
\end{equation}
For a Milky Way-like galaxy with $v_\mathrm{c} = 200$ km s$^{-1}$ and a cluster at $r_\mathrm{g} = 3$ kpc, this gravitational force is equal to the force from the bubble pressure when the shock is at a radius
\begin{align}
	R_\mathrm{max} &= \left( \frac{9 (\gamma-1)}{4 \pi} \frac{1}{\rho_0} \frac{v_\mathrm{c}^2}{r_\mathrm{g}^2} E_\mathrm{R, int} \right)^{1/5}
	\\
	&\approx 700 \left( \frac{N_\mathrm{SNe}}{1000} \right)^{1/5} \left( \frac{\rho_0}{1.33 \times m_\mathrm{H} \text{ cm}^{-3}} \right)^{1/5}  \nonumber
	\\
	& \quad\quad \times \left( \frac{v_\mathrm{c}}{200 \text{ km s}^{-1}} \right)^{-2/5}   \left( \frac{r_\mathrm{g}}{3 \text{ kpc}} \right)^{2/5}   \text{ pc}.
\end{align}
As with our self-gravity model, we create a model which truncates our simulations if and when they reach this radius. In \autoref{fig:gravity:disc}, we compare the results of this galactic gravity model to a subset of our simulations. Similar to the self-gravity results shown in \autoref{fig:gravity:self}, galactic gravity has the largest effect for the largest clusters. We also find that this galactic gravity model is able to reduce the momentum by a greater factor than our self-gravity model; the self-gravity model never reduced the momentum by more than a factor of 1.4, whereas the galactic gravity model can reduce the momentum by up to a factor of 3.3. As with self-gravity, however, we caution that for many applications, the momentum budget of interest will be that excluding the effects of disc gravity, since if disc gravity is included in the overall model, it should not be double-counted by also including it in the subgrid feedback model.

\begin{figure}
\includegraphics[width=\columnwidth]{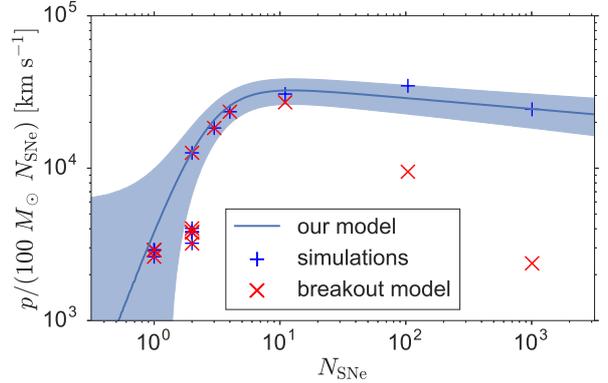}
\caption{
Comparison of simulations with no disc breakout (marked by a blue $+$) and our simple post-processing model of disc breakout for all of our simulations with $Z=Z_\odot$ and $\rho = 1.33 \times 10^{-3}$ $m_\mathrm{H}$ cm$^{-3}$.
}
\label{fig:gravity:breakout}
\end{figure}

Our disc gravity model is incomplete; among other things, it ignores the density and pressure gradients that result from this gravitational potential.  As a superbubble expands vertically, it will find less dense and lower pressure gas; as it expands horizontally, it will not experience such gradients in the background gas.  This will break the spherical symmetry of the bubble expansion, and can lead to disc breakout where the bubble expansion becomes predominantly vertical.  When studying this phenomenon with hydrodynamic simulations, \citet{1989ApJ...337..141M} found that spherical symmetry is often broken and shell mixing instabilities are significant by the time the bubble has expanded 3-4 scale heights in the vertical direction. Applying this to a Milky Way-like galaxy, using a thin-disc scale height of 100 pc, we can create a model which cuts off the momentum growth when a bubble reaches $R_\mathrm{max} = 400 $ pc.  
The results of this model can be seen in \autoref{fig:gravity:breakout}. As with our previous models, the effect of this model is stronger for larger clusters, but this model predicts effects which increase much more rapidly as cluster size increases. While there is no significant effect on our 11 SNe simulation, it has the largest effect for the 1008 SNe simulation for all of the models we have tested, decreasing the momentum by a factor of about 10.  It is important to understand that this behaviour is not a result of the total momentum decreasing for large bubbles. It is simply that, for the largest clusters, the bubble expands to the breakout radius before a significant fraction of the SNe occur, and, by assumption, these additional SNe then contribute no additional momentum. This causes the average momentum per SN to fall, because the extra SNe are counted in the denominator but not the numerator of our average.

\begin{figure}
\includegraphics[width=\columnwidth]{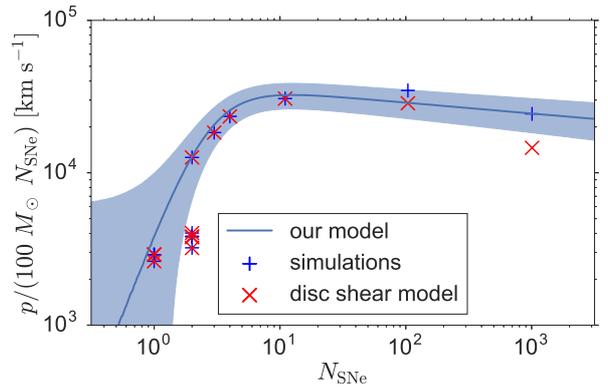}
\caption{
Comparison of simulations with no rotational shear (marked by a blue $+$) and our simple post-processing model of rotational shear in a disc for all of our simulations with $Z=Z_\odot$ and $\rho = 1.33 \times 10^{-3}$ $m_\mathrm{H}$ cm$^{-3}$.
}
\label{fig:gravity:shear}
\end{figure}

In addition to considering distortions caused by expansion perpendicular to the disc plane, we can also consider distortions due to shear within the disc plane. Adopting a shear timescale
\begin{equation}
	t_\mathrm{shear} = \Omega_\mathrm{orb}^{-1} \frac{r_\mathrm{g}}{R_\mathrm{shell}}
\end{equation}
for a disc orbital frequency $\Omega_\mathrm{orb} = v_\mathrm{c} / r_\mathrm{g}$, we can truncate our bubble evolution when the bubble age ($t - t_\text{first SN}$) exceeds the shear time. Results can be seen in \autoref{fig:gravity:shear}. Once again, only the largest clusters are significantly affected, with the largest effect being a decrease of a factor of 1.75 for the 1008 SNe cluster.

Comparing these post-processing models, we find that all of the galactic models (gravity, breakout and shear) are predicted to have larger effects than the self-gravity model, but these galactic models introduce a number of free parameters which we have not explored (disc circular velocity, galactocentric radius and disc scale height).  It is important to remember that these models are only very rough estimates. When investigating self-gravity, we found that directly including self-gravity in simulations led to a much stronger effect than predicted by our simple model.  The same is likely true for these galactic effects, which we cannot directly include in our simulations. Still, even though every model was structured to decrease the final momentum, we always saw that clustering can lead to an increased momentum efficiency, compared to our single SN results.  Moreover, all of these corrections are relatively small for the most efficient clusters, those producing $\sim 10$ SNe ($M_\mathrm{cluster} \sim 1000 M_\odot$).

Overall, this suggests that superbubble models are less cleanly separable from galactic dynamics than single SNR models.  Single, isolated SNRs might only expand to 100 pc over 2 Myr, allowing us to largely ignore effects like disc shear and galactic gravity. As we have shown, these effects can play a significant role for models of clustered SNe and superbubbles.  Testing these effects self-consistently goes beyond the capability of our 1D simulations, but we have already gained some insight from our simplified post-processing models.  By design, all of these models lowered the final momenta of our bubbles (by as much as a factor of 10, for the disc breakout model applied to our largest bubble), but every model still predicted that clustered SNe lead to enhanced momentum feedback (for instance, our disc breakout model still predicts an average momentum per SN 4 times larger than the single SN value).  Both clustering of SNe and effects from the host galaxy seem to play significant roles in the overall SNe momentum budget; it would be useful for higher dimensional simulations to explore these two effects simultaneously in the future.

\section{Conclusions} 
\label{section:conclusion}

We perform several hundred 1D simulations to study the momentum delivered to the interstellar medium by clustered supernova explosions over a wide range of star cluster sizes, gas densities and metallicities.  Our simulations use a realistic IMF paired with realistic stellar lifetimes, and we evolve them at very high numerical resolution until the momentum of the expanding shell reaches a maximum. At the end of our simulations, we find that our clusters typically retain $1-10$\% of the injected SN energy (similar to isolated SNe), but clustered SNe produce significantly more momentum per SN than isolated SNe, i.e., the momentum injected by a star cluster is a superlinear function of the number of SNe that explode within it.  Clustering has the largest impact for $10-100$ SNe, leading to an order of magnitude increase in the momentum per SN.  When integrating over the observed cluster mass function, our findings suggest that the mean SN momentum budget per mass of stars formed is $p/m_* \approx 1-2 \times 10^4$ km s$^{-1}$, which is $\sim 0.5-1$ dex higher than the value of $\approx 3000$ km s$^{-1}$ most commonly adopted in the literature.

The increased momentum budget for SNe will have strong implications for any simulation or analytic model that relies on SN momentum injection. In galaxies where the overall star formation rate is high enough for clustering of SNe to be significant for quenching star formation and producing galactic winds (i.e., perhaps not in dwarfs, but almost certainly in more massive galaxies), using our updated SN momentum budget may cause these models' predicted star formation rates to decrease by the same factor of $\sim 0.5-1$ dex by which the SN momentum budget increases. This may render the models inconsistent with the observed relationship between gas and star formation rate, which will require changes in other free parameters to bring the models back into agreement.  The implications for galactic wind launching are less clear, since the increased SN momentum budget will be offset by overall lower star formation rates.

To facilitate the implementation of our results in 3D numerical simulations that include explicit supernova momentum injection, we provide a fitting formula for the momentum per SN as a function of cluster size, ambient density and metallicity. This formula is suitable for implementation in  galaxy simulations capable of resolving individual star clusters, typically $\sim 10^2 - 10^5$ $M_\odot$. We also provide tabulated outputs from our simulations, for those who wish to calibrate subgrid models at a range of size scales. For lower resolution simulations we recommend the value of $ p / m_* \approx 1-2 \times 10^4$ km s$^{-1}$ we obtain by integrating over the cluster mass function.  Regardless of resolution, it is clear that the existing body of simulations may need to be revisited, and some of the strong assumptions that previous authors have adopted to make feedback more effective (e.g., assuming efficient trapping of infrared radiation) may be relaxed. Other numerical schemes, such as turning off radiative cooling for an extended period of time \citep[e.g.,][]{2006MNRAS.373.1074S} or stochastically injecting thermal energy in order to delay cooling \citep[e.g.,][]{2012MNRAS.426..140D}, may prove to be closer to reality than had previously been assumed.

Properly capturing the effect of clustering in a 1D simulation requires very high numerical resolution to avoid over-cooling through numerical mixing. The resolution requirements may be less severe, and the momentum injection rate lower, in higher dimensions where instabilities may produce mixing at large physical scales. However, the size scale and mixing rate due to instabilities likely depends strongly on the properties of the host galaxy, the nature of the background into which the SNR is propagating and the microphysical details like magnetic fields and conduction near the contact discontinuity. Since we are unable to directly include these effects in our simulations, we create simple models to estimate the strengths of some of these effects. While these models are able to decrease the momentum of the most-massive cluster by a factor of 10, lower mass (more common) clusters are less affected. For each of our models, the average momentum per SN remains greater than the fiducial, unclustered value by a factor of 4, assuming a realistic cluster mass distribution.
But these are just simple estimates; no present simulation includes all these effects, and thus the correct momentum budget for clustered SNe in multiple dimensions remains uncertain. While the correct momentum budget may not be as high as $\sim 0.5 - 1$ dex greater than the commonly-adopted single SN value (as we find in one dimension), it is still likely greater than the single SN value. There is clearly an urgent need for further study.

\section*{Acknowledgements}

We thank Justin Brown, John Forbes, Anna Rosen and Brant Robertson for useful discussions. We thank the referee, Norm Murray, for a constructive and insightful report.  This work was supported by the NSF through grants AST-1405962 (ESG, MRK and AD), AST-1229745 (PM) and DGE 1339067 (ESG), by the Australian Research Council through grant ARC DP160100695 (MRK) and by NASA through grant NNX12AF87G (PM). AD's work is also supported by the grants ISF 124/12, I-CORE Program of the PBC/ISF 1829/12, and BSF 2014-273. MRK thanks the Simons Foundation, which contributed to this work through its support for the Simons Symposium ``Galactic Superwinds: Beyond Phenomenology''. PM thanks the Pr\'{e}fecture of the Ile-de-France Region for the award of a Blaise Pascal International Research Chair, managed by the Fondation de l'Ecole Normale Sup\'{e}rieure.



\bibliography{SNe.bbl}


\appendix

\section{Code Verification}
\label{section:verification}

\subsection{Sedov Verification}
\label{section:verification:Sedov}

The mass and energy of each supernova is injected into the innermost zone, with all the energy injected as thermal energy.  This is not a realistic configuration; at no stage do we expect a uniformly mixed sphere, on the order of $0.1$ parsecs in radius, which is over-pressured but not yet expanding.  Given these convenient but unphysical initial conditions, we need to verify that our system will evolve into a realistic configuration.

\begin{figure}
\includegraphics[width=\columnwidth]{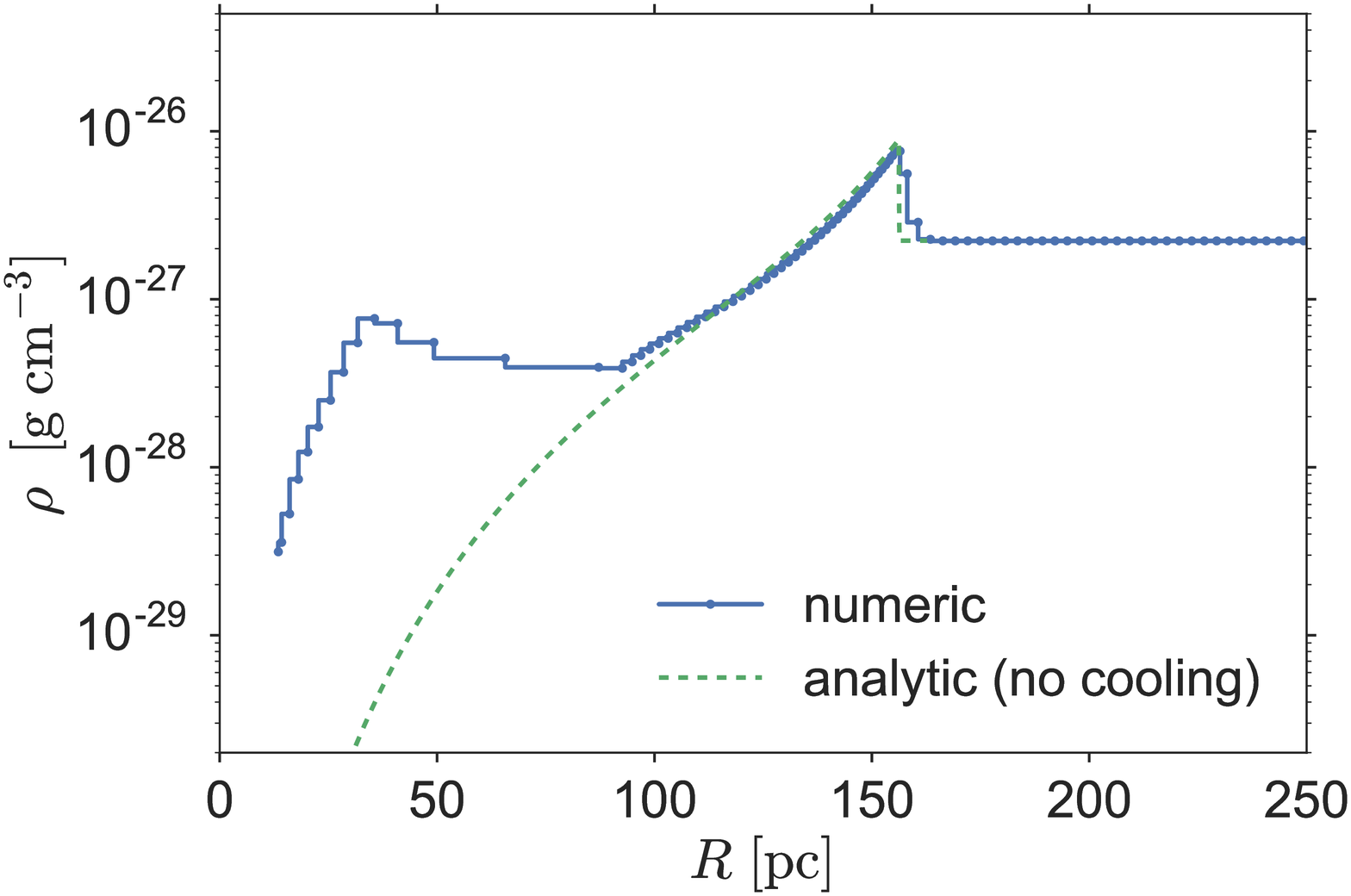}
\caption{
Comparison of our numeric results (solid line) against the analytic Sedov solution (dashed line) for a $\rho = 1.33 \times 10^{-3}$ $m_\mathrm{H}$ cm$^{-3}$, $Z=Z_\odot$, $M_\mathrm{cluster}=10^2$ $M_\odot$ ($N_\mathrm{SNe}=1$) cluster, at $t=.17$ Myr.
}
\label{fig:verification:Sedov}
\end{figure}

We can look at an early time snapshot of a single SN simulation to verify that the system accurately relaxes into a physical configuration.  At early times, cooling losses should still be negligible, so we expect our system to be in the Sedov phase.  \autoref{fig:verification:Sedov} shows a snapshot of our numerical results, compared with the analytic Sedov prediction.  There is a noticeable overdensity at inner radii, but this is to be expected: the analytic Sedov solution assumes no ejecta mass, whereas our simulation includes ejecta mass.  The extra ejecta mass appears as an overdensity at inner radii. Excepting that, our simulation is in good agreement with the Sedov prediction so we consider our injection scheme valid.

\subsection{Thornton et al. Verification}
\label{section:verification:Thornton} 

\begin{figure}
\includegraphics[width=\columnwidth]{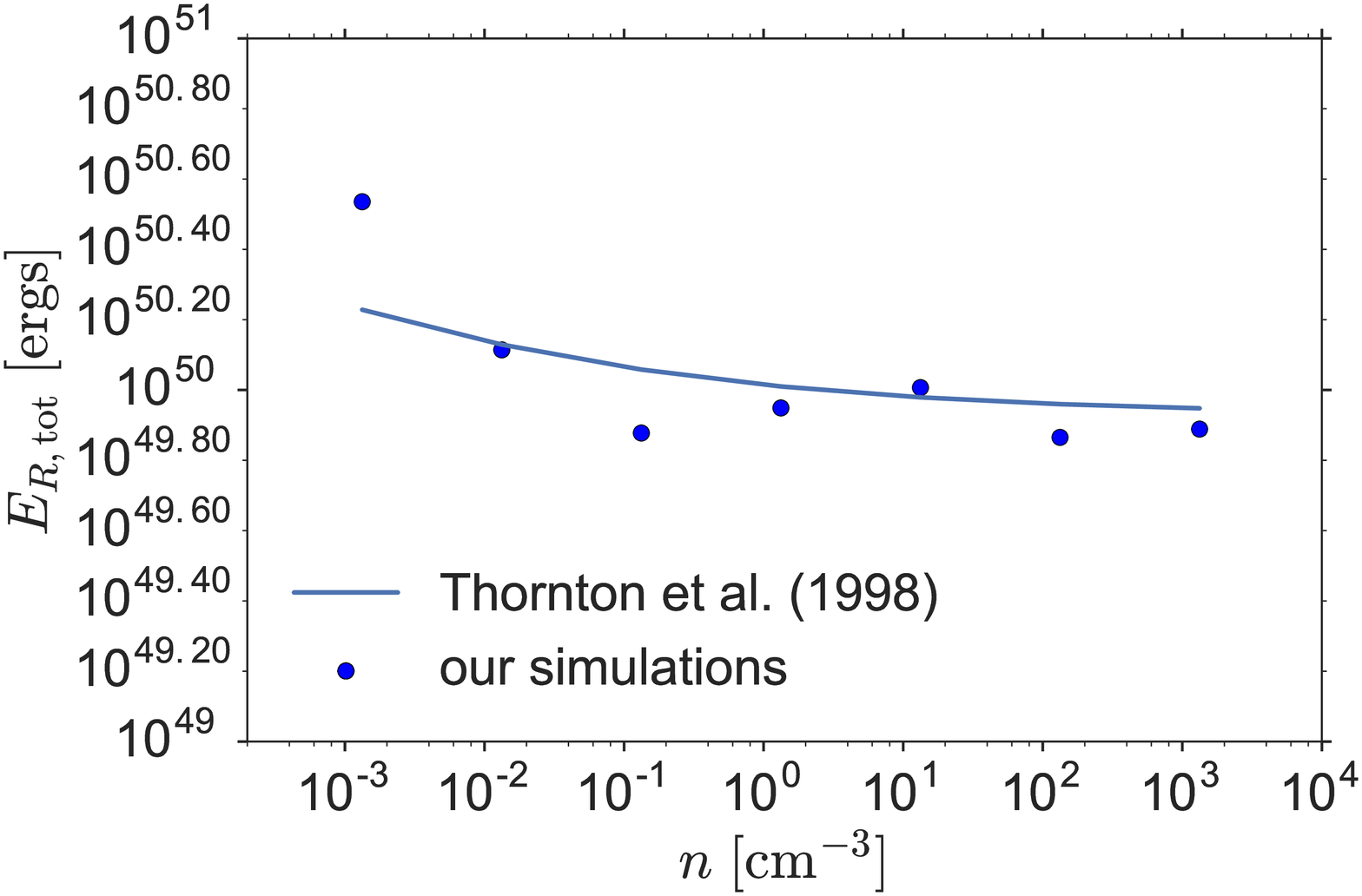}
\caption{
Verification that our code can reproduce the results of \citet{1998ApJ...500...95T}, for total energy contained within the SNR ($E_\mathrm{R,tot}$) at the completion time defined by \citet{1998ApJ...500...95T}.
}
\label{fig:verification:Thornton:density}
\end{figure}

\begin{figure}
\includegraphics[width=\columnwidth]{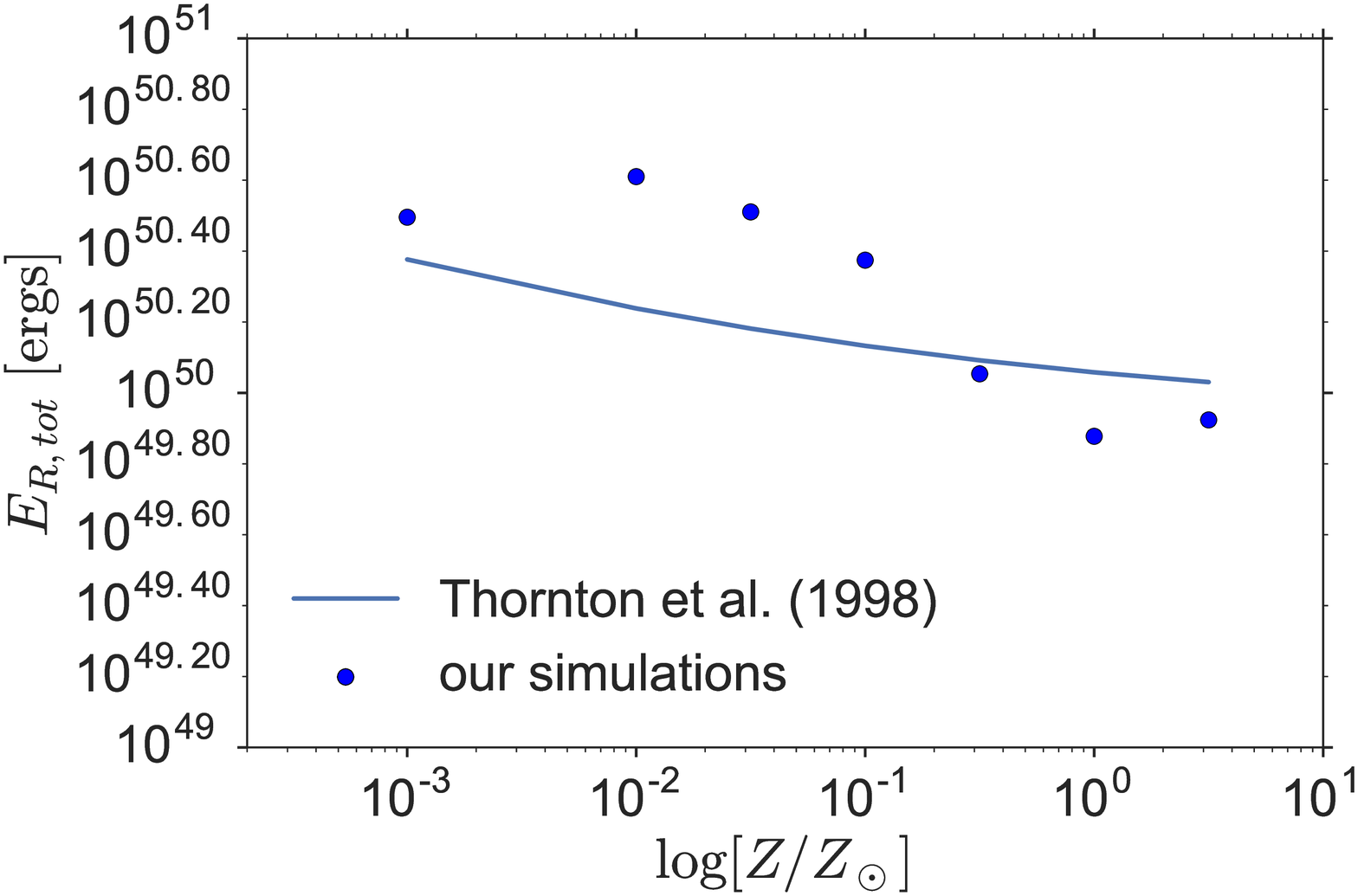}
\caption{
Same as \autoref{fig:verification:Thornton:density}, but now with total energy as a function of metallicity.
}
\label{fig:verification:Thornton:metallicity}
\end{figure}

We also verified our code against the results of \citet{1998ApJ...500...95T}, who measured the total energy from single SNe.  We ran single SN simulations at the same background conditions as \citeauthor{1998ApJ...500...95T}, fixing the SN ejecta mass to be $3$ $M_\odot$ with an ejecta metallicity equal to the background metallicity, and extracting results at the same time as \citeauthor{1998ApJ...500...95T}.  We compare our simulations to the model provided by \citeauthor{1998ApJ...500...95T} in \autoref{fig:verification:Thornton:density} and \autoref{fig:verification:Thornton:metallicity}.

We judge that our residuals are comparable to the residuals present in the data of \citeauthor{1998ApJ...500...95T}, and we are not surprised that there are discrepancies. We use different initial conditions: our simulation injects all of the SNe energy into the innermost zone as thermal energy, while \citeauthor{1998ApJ...500...95T} spreads the energy across 150 zones, and adds some of it as kinetic energy. We were able to use different initial conditions because we used different hydrodynamic solvers: \citeauthor{1998ApJ...500...95T} uses a finite-difference method which cannot handle the strong shock that occurs by injecting all the energy into one zone, while our finite-volume method is much more robust to these strong shock conditions.  Finally, we use a cooling package that differs from the cooling function used by \citeauthor{1998ApJ...500...95T}.  All these differences lead us to expect the minor discrepancies between our results and the results of \citeauthor{1998ApJ...500...95T}.

\bsp	
\label{lastpage}
\end{document}

%% file: results.stub.table.tex
\begin{table*}
\caption{Overview of Numeric Results. The table shown here is only a stub; it is provided in its entirety as a Machine-Readable Table. The Machine-Readable Table also includes a column with an id unique to each row, to allow cross-referencing with \autoref{tab:evolution}, which has been hidden here to save space. }
\label{tab:results}
\begin{tabular}{cccrrrccccc}
$\rho$ & $Z$ & $M_\mathrm{cluster}$ & $N_\mathrm{SNe}$ &  $t$ & $R$ & $M_\mathrm{R}$ & $p$  & $E_\mathrm{R,kin}$ & $E_\mathrm{R,int}$ & flag\\
(1.33 $m_\mathrm{H}$ cm$^{-3}$) & ($Z_\odot$) & ($M_\odot$) &  & (Myr) & (pc) & ($M_\odot$) & (g cm s$^{-1}$) & (ergs) & (ergs) &  \\
\hline
$10^{+0}$  & $ 10^{+0.0}$  & $10^{2.0}$  & 1  & $5.9$  & $56.0$  & $ 2.42 \times 10^{4}$ & $ 5.22 \times 10^{43}$ & $ 3.14 \times 10^{49}$ & $ 1.77 \times 10^{48}$  & 0 \\
$10^{+0}$  & $ 10^{+0.0}$  & $10^{2.5}$  & 3  & $55.3$  & $294.3$  & $ 3.51 \times 10^{6}$ & $ 1.10 \times 10^{45}$ & $ 9.47 \times 10^{49}$ & $ 2.73 \times 10^{50}$  & 0 \\
$10^{+0}$  & $ 10^{+0.0}$  & $10^{3.0}$  & 11  & $91.8$  & $533.0$  & $ 2.08 \times 10^{7}$ & $ 6.72 \times 10^{45}$ & $ 5.94 \times 10^{50}$ & $ 1.63 \times 10^{51}$  & 0 \\
$10^{+0}$  & $ 10^{+0.0}$  & $10^{4.0}$  & 104  & $173.3$  & $1149.0$  & $ 2.09 \times 10^{8}$ & $ 7.18 \times 10^{46}$ & $ 6.67 \times 10^{51}$ & $ 1.63 \times 10^{52}$  & 0 \\
$10^{+0}$  & $ 10^{+0.0}$  & $10^{5.0}$  & 1008  & $285.2$  & $2133.0$  & $ 1.34 \times 10^{9}$ & $ 4.88 \times 10^{47}$ & $ 4.79 \times 10^{52}$ & $ 1.04 \times 10^{53}$  & 0 \\
\end{tabular}
\end{table*}

%% file: evolution.stub.table.tex
\begin{table*}
\caption{ Momentum Evolution. The table shown here is only a stub; it is provided in its entirety as a Machine-Readable Table.}
\label{tab:evolution}
\begin{tabular}{crrccccrccc}
ID & $t$ & $R_\mathrm{shock}(t)$  & $M_\mathrm{R}(t)$ & $p(t)$ & $E_\mathrm{R, kin}(t) $ &  $E_\mathrm{R, int}(t)$  & $N_\mathrm{SNe}(< t)$ \\
 & (Myr) & (pc) & ($M_\odot$) & (g cm s$^{-1}$) & (ergs) & (ergs)  &  \\
\hline
25451948-485f-46fe-b87b-f4329d03b203  & $4.0$  & $1.3$  & $ 6.92 \times 10^{0}$ & $ 0.00 \times 10^{0}$ & $ 0.00 \times 10^{0}$ & $ 1.00 \times 10^{51}$ & $1$ \\
25451948-485f-46fe-b87b-f4329d03b203  & $5.0$  & $61.4$  & $ 3.20 \times 10^{4}$ & $ 1.73 \times 10^{44}$ & $ 2.58 \times 10^{50}$ & $ 5.84 \times 10^{50}$ & $2$ \\
25451948-485f-46fe-b87b-f4329d03b203  & $10.2$  & $152.9$  & $ 4.92 \times 10^{5}$ & $ 1.24 \times 10^{45}$ & $ 8.26 \times 10^{50}$ & $ 1.14 \times 10^{51}$ & $5$ \\
25451948-485f-46fe-b87b-f4329d03b203  & $15.4$  & $209.2$  & $ 1.26 \times 10^{6}$ & $ 2.18 \times 10^{45}$ & $ 9.88 \times 10^{50}$ & $ 1.40 \times 10^{51}$ & $7$ \\
25451948-485f-46fe-b87b-f4329d03b203  & $20.5$  & $249.3$  & $ 2.13 \times 10^{6}$ & $ 2.83 \times 10^{45}$ & $ 1.07 \times 10^{51}$ & $ 2.43 \times 10^{51}$ & $9$ \\
25451948-485f-46fe-b87b-f4329d03b203  & $25.6$  & $283.1$  & $ 3.12 \times 10^{6}$ & $ 3.56 \times 10^{45}$ & $ 1.06 \times 10^{51}$ & $ 1.61 \times 10^{51}$ & $9$ \\
25451948-485f-46fe-b87b-f4329d03b203  & $35.3$  & $335.1$  & $ 5.18 \times 10^{6}$ & $ 4.59 \times 10^{45}$ & $ 1.05 \times 10^{51}$ & $ 2.84 \times 10^{51}$ & $11$ \\
\end{tabular}
\end{table*}

%% file: bayesian_fit_intervals.tex
            \begin{align}
                \left( \frac{p}{N_\mathrm{SNe}} \right)_{0,\mathrm{few}} &= 4249^{+741}_{-683} \cdot 100 \; M_\odot \text{ km s}^{-1}
                \\
                \eta_{Z,\mathrm{few}} &= 0.05^{+0.05}_{-0.06}
                \\
                \eta_{\rho,\mathrm{few}} &= -0.06^{+0.03}_{-0.03}
                \\
                 \eta_{N,\mathrm{few}} &= 2.20^{+0.24}_{-0.23}
                \\
                \left( \frac{p}{N_\mathrm{SNe}} \right)_{0,\mathrm{many}} &= 23546^{+1072}_{-1073} \cdot 100 \; M_\odot \text{ km s}^{-1}
                \\
                \eta_{Z,\mathrm{many}} &= 0.15^{+0.01}_{-0.01}
                \\
                \eta_{\rho,\mathrm{many}} &= 0.14^{+0.01}_{-0.01}
                \\
                \eta_{N,\mathrm{many}} &= -0.07^{+0.02}_{-0.02}
                \\
                \sigma &= 6075^{+214}_{-202} \cdot 100 \; M_\odot \text{ km s}^{-1}
            \end{align}